\documentclass[twocolumn,twocolappendix,showpacs,superscriptaddress,groupedaddress,
  nofootinbib,floatfix]{aastex63}

\pdfoutput=1 
\usepackage{amsmath,amsfonts,amstext,amssymb}
\usepackage[T1]{fontenc}
\usepackage[utf8]{inputenc}
\usepackage{graphicx}
\usepackage{rotating}
\usepackage{hyperref}
\usepackage{xspace}
\usepackage{enumerate}

\usepackage{xcolor}

\newcommand{\be}{\begin{eqnarray}}
\newcommand{\ee}{\end{eqnarray}}

\newcommand{\tGRAthena}{\texttt{GR-Athena++}}
\newcommand{\GRAthena}{\tGRAthena\xspace}

\newcommand{\tAthena}{\texttt{Athena++}}
\newcommand{\Athena}{\tAthena\xspace}

\newcommand{\tBAM}{\texttt{BAM}}
\newcommand{\BAM}{\tBAM\xspace}

\newcommand{\treprimand}{\texttt{RePrimAnd}}
\newcommand{\reprimand}{\treprimand\xspace}
\newcommand{\tPrimitiveSolver}{\texttt{PrimitiveSolver}}
\newcommand{\PrimitiveSolver}{\tPrimitiveSolver\xspace}

\def\p{\partial}

\def\d{{\rm d}}
\def\e{{\rm e}}
\def\i{{\rm i}}
\def\f{{\rm f,~BH}}
\def\gccm{{\rm g\,cm^{-3}}}
\def\Msun{{\rm M_{\odot}}}

\def\GMc2{{\rm G M_{\odot} c^{-2}}}

\def\eps{\epsilon}
\def\eps{\epsilon}
\def\B{\mathcal{B}}
\def\I{\mathcal{I}}
\def\M{\mathcal{M}}
\def\O{\mathcal{O}}
\def\R{\mathcal{R}}

\def\l{\ell}
\def\nn{\nonumber}

\def\kt2{\kappa^\text{T}_2}

\def\Mo{{\rm M_{\odot}}}
\def\kt2{\kappa^\text{T}_2}

\def\half{\frac{1}{2}}

\def\D{\mathrm{D}}
\def\pd{\partial{}}

\def\2nd{2^\mathrm{nd}}
\def\4th{4^\mathrm{th}}
\def\6th{6^\mathrm{th}}
\def\8th{8^\mathrm{th}}

\def\eg{\textit{e.g.}}
\def\ie{\textit{i.e.}}

\newcommand{\mc}[1]{\ensuremath{\mathcal{#1}}}

\def\z4c{$\mathrm{Z}4\mathrm{c}$}
\def\z4oc{$\mathrm{Z}4(\mathrm{c})$}
\def\z4{$\mathrm{Z}4$}
\def\ccz4{$\mathrm{CCZ}4$}

\newcommand{\Mesh}{\texttt{Mesh}}
\newcommand{\MeshBlock}{\texttt{MeshBlock}}

\usepackage{color}
\definecolor{cyan}{rgb}{0,0.9,0.9}
\definecolor{orange}{rgb}{0.9,0.5,0}
\definecolor{magenta}{rgb}{1,0,1}
\definecolor{purple}{rgb}{0.8,0.4,0.8}
\definecolor{gray}{rgb}{0.5,0.5,0.5}

\begin{document}

\title{\GRAthena\,: General-relativistic magnetohydrodynamics simulations of neutron star spacetimes}
\author[0000-0003-2244-3462]{William \surname{Cook}}
\affiliation{
  Theoretisch-Physikalisches Institut, Friedrich-Schiller-Universit{\"a}t Jena, 07743, Jena, Germany}
\author{Boris \surname{Daszuta}}
\affiliation{
  Theoretisch-Physikalisches Institut, Friedrich-Schiller-Universit{\"a}t Jena, 07743, Jena, Germany}
\author[0000-0001-5705-1712]{Jacob \surname{Fields}}
\affiliation{Institute for Gravitation and the Cosmos, The Pennsylvania State University,
  University Park, PA 16802, USA}
\affiliation{Department of Physics, The Pennsylvania State University, University Park,
  PA 16802, USA}
\author{Peter \surname{Hammond}}
\affiliation{Institute for Gravitation and the Cosmos, The Pennsylvania State University,
  University Park, PA 16802, USA}
\author[0000-0001-7345-4415]{Simone \surname{Albanesi}}
\affiliation{Dipartimento di Fisica, Università di Torino, Torino, 10125, Italy}
\affiliation{INFN sezione di Torino, Torino, 10125, Italy}
\author{Francesco \surname{Zappa}}
\affiliation{
  Theoretisch-Physikalisches Institut, Friedrich-Schiller-Universit{\"a}t Jena, 07743, Jena, Germany}
\author[0000-0002-2334-0935]{Sebastiano \surname{Bernuzzi}}
\affiliation{
  Theoretisch-Physikalisches Institut, Friedrich-Schiller-Universit{\"a}t Jena, 07743, Jena, Germany}
\author[0000-0001-6982-1008]{David \surname{Radice}}
\thanks{Alfred P.~Sloan Fellow}
\affiliation{Institute for Gravitation and the Cosmos, The Pennsylvania State University, University Park, PA 16802, USA}
\affiliation{Department of Physics, The Pennsylvania State University, University Park, PA 16802, USA}
\affiliation{Department of Astronomy and Astrophysics, The Pennsylvania State University, University Park, PA 16802, USA}
\date{\today}

\begin{abstract}
We present the extension of \GRAthena{} to general-relativistic
magnetohydrodynamics (GRMHD) for applications to
neutron star spacetimes.
The new solver couples the constrained transport implementation of \Athena{} to the Z4c
formulation of the Einstein equations to simulate dynamical spacetimes
with GRMHD using oct-tree adaptive mesh refinement. 
We consider benchmark problems for isolated and binary
neutron star spacetimes demonstrating stable and convergent results at
relatively low resolutions and without grid symmetries imposed. 
The code correctly captures magnetic field
instabilities in non-rotating stars with total relative violation of the divergence-free
constraint of $10^{-16}$. It handles evolutions with a
microphysical equation of state and black hole formation in the
gravitational collapse of a rapidly rotating star.  
For binaries, we demonstrate correctness of the evolution under
the gravitational radiation reaction and show convergence of gravitational
waveforms. We showcase the use of adaptive mesh refinement to resolve
the Kelvin-Helmholtz instability at the collisional interface in a merger of magnetised binary neutron stars.  
\GRAthena{} shows strong scaling efficiencies above $80\%$ in excess of
 $ 10^5$ CPU cores and excellent weak scaling is
shown up to $\sim 5 \times 10^5$ CPU cores in a realistic production setup.
\GRAthena{} allows for the robust simulation of GRMHD flows in
strong and dynamical gravity with exascale computers. 
\end{abstract}

\section{Introduction}\label{sec:intro}

The detection of the gravitational wave (GW) signal GW170817,
 combined with the associated observation of
the electromagnetic (EM) counterpart, 
a short Gamma Ray burst (SGRB),  GRB 170817A, alongside a kilonova, from the merger of a Binary Neutron Star (BNS) system
marked the beginning of the era of multimessenger astronomy \citep{TheLIGOScientific:2017qsa,Goldstein:2017mmi,Savchenko:2017ffs}. The near simultaneous detections of these two 
signals confirmed BNSs as progenitors of SGRBs, and provided an insight 
both into fundamental physics questions within General Relativity (GR) and the astrophysical
origins of such high energy phenomena \citep{Monitor:2017mdv}.

The end products of BNS mergers have long been viewed as candidates for the 
launching of relativistic jets which give rise to SGRBs such as GRB 170817A 
\citep{Blinnikov:1984a,Paczynski:1986px,Goodman:1986a,Eichler:1989ve,Narayan:1992iy}, though the 
precise mechanism through which these jets are launched is still an open question. 
The presence of large magnetic 
fields in the post merger remnant, either in a highly magnetised magnetar neutron star (NS) or in the vicinity of an accreting 
Black Hole (BH), may however play a role in the formation of the jet
, see \eg~\citep{Piran:2004ba,Kumar:2014upa,Ciolfi:2018tal} for reviews. 
Such large fields can be produced through amplifications arising as the result of magnetohydrodynamic
(MHD) instabilities during the inspiral such as the Kelvin-Helmholtz instability (KHI) 
\citep{Rasio:1999a,Price:2006fi,Kiuchi:2015sga}, magnetic winding, and the Magneto-Rotational 
Instability (MRI) \citep{Balbus:1991a}, which also lead to a rearrangement of the initial magnetic 
field structure, possibly driving turbulence in the remnant disc.

Even before the merger of two magnetised NSs however, certain magnetic field configurations within an isolated NS 
are known to be susceptible to various instabilities, specifically in the case of an initially
poloidal magnetic field 
\citep{Tayler:1957a,Tayler:1973a,Wright:1973a,Markey:1973a,Markey:1974a,Flowers:1977a}. 
Consequently the early time configuration of magnetic fields within a BNS system during
the inspiral phase is also an open question, with long-term simulations of isolated stars in static and dynamical
spacetimes suggesting that higher resolution is required to understand reasonable initial
configurations for the magnetic fields 
\citep{Kiuchi:2008ss,Ciolfi:2011xa,Ciolfi:2013dta,Lasky:2011un,Pili:2014npa,Pili:2017yxd,Sur:2021awe}, 
while thermal effects, such as those summarised in \cite{Pons:2019zyc}, also  play a key role in the evolution of the field.

As well as driving the production of SGRBs, the distribution and nature of matter outflowing
from a BNS merger may be influenced by the presence of magnetic fields during the merger 
\citep{Siegel:2014ita,Kiuchi:2014hja,Siegel:2017nub,Mosta:2020hlh,Curtis:2021guz,Combi:2022nhg,deHaas:2022ytm,Kiuchi:2022nin,Combi:2023yav}.
These outflows determine the nature of the long lived EM signal that follows the BNS merger, such
as AT2017gfo which accompanied GW170817 \citep{GBM:2017lvd,Coulter:2017wya,Soares-santos:2017lru,Arcavi:2017xiz}, 
known as the kilonova \citep{Li:1998bw,Kulkarni:2005jw,Metzger:2010sy},
as well as the r-process nucleosynthesis responsible for heavy element production that
occurs in the neutron rich matter that is ejected from the system \citep{Pian:2017gtc,Kasen:2017sxr}.

To approach a full understanding of a BNS system from its late inspiral, 
through merger, to the post-merger evolution; we must model a broad range of 
physical processes, including GR, General 
Relativistic Magnetohydrodynamics (GRMHD), weak nuclear processes leading to 
neutrino emission and reabsorption, and the finite temperature behaviour of 
dense nuclear matter. 
Fully modelling such a challenging problem on varying length and time
scales requires the use of a numerical approach with adaptive mesh refinement (AMR). 

Such simulations of BNS spacetimes with numerical relativity codes evolving the Einstein and Euler 
equations, increasingly in combination with Maxwell's equations in the ideal MHD approximation, 
have been extensively performed over the last 25 years, with a wide variety of codes 
developed for this purpose over this period, \citep{Shibata:1999hn,Font:1998hf,Shibata:1999wm,
Duez:2002bn,Baiotti:2004wn,Duez:2005sf,Shibata:2005gp,Anderson:2006ay,Anderson:2007kz,
Giacomazzo:2007ti,Duez:2008rb,Etienne:2010ui,Liebling:2010bn,Foucart:2010eq,Thierfelder:2011yi,East:2011aa,
Loffler:2011ay,Moesta:2013dna,Radice:2013xpa,Etienne:2015cea,
Kidder:2016hev,Palenzuela:2018sly,Vigano:2018lrv,
Cipolletta:2019geh,Cheong:2020kpv,Rosswog:2020kwm,Shankar:2022ful}. 
These codes couple free evolution schemes for the Einstein equations to GRMHD evolutions.

For the solution of the Einstein equations common modern approaches are given by the BSSN \citep{Shibata:1995we,Baumgarte:1998te}, 
Z4c \citep{Bernuzzi:2009ex,Ruiz:2010qj,Weyhausen:2011cg,Hilditch:2012fp}, and CCZ4  formulations 
\citep{Alic:2011gg,Alic:2013xsa}, with moving puncture gauge conditions 
\citep{Brandt:1997tf,Campanelli:2005dd,Baker:2005vv}; and the Generalised Harmonic Gauge approach
\citep{Pretorius:2004jg,Pretorius:2005gq}.
The equations of GRMHD are generally written in a conservative
formulation,
where conservative schemes are key to ensure that shocks
are correctly captured and mass is conserved
\citep{Anile:1990a,Marti:1991wi,Banyuls:1997zz,Balsara:1998b,Komissarov:1999a,Gammie:2003rj,Anninos:2005kc,Komissarov:2005wj,Anton:2005gi,DelZanna:2007pk}, with a thorough review provided by \cite{Font:2007zz}.

The numerical scheme used to ensure that the divergence-free condition on the magnetic 
field is preserved varies also between codes. Schemes employed include
the Constrained Transport (CT) algorithm 
of \cite{Evans:1988a,Ryu:1998ar,Londrillo:2003qi} with face centred magnetic field 
discretisation; the Flux-CT method \citep{Toth:2000}, with a cell centred magnetic 
field discretisation; casting the equations in terms of the vector potential $A$, \eg~\cite{Etienne:2011re}
; and divergence cleaning, where divergence violations are propagated and damped through a
  hyperbolic equation~\citep{Dedner:2002a}.

In this paper we demonstrate the ability of the code \GRAthena to evolve 
GRMHD problems in dynamically evolving spacetimes for the first time. \GRAthena is built on top of 
the GRMHD code \Athena \citep{Stone:2020}, removing its restriction to stationary spacetimes, 
with the evolution of the Einstein Equations detailed in \cite{Daszuta:2021ecf},
allowing the evolution of dynamical problems such as the merger of magnetised BNSs. This code
allows us to exploit the infrastructure and numerical schemes implemented in \Athena designed for 
accurate evolution of the GRMHD system, with minimal violation of the divergence-free condition 
on the magnetic field.

\Athena uses a block-based, oct-tree AMR structure, which
forgoes the need for synchronisation calls required in the Berger-Oliger time subcycling algorithm
present in codes with AMR structures featuring nested grids, as well as task-based parallelism
for executing the evolution loop. This allows \Athena to scale on up to $\mathcal{O}(10^5)$  CPU 
cores, allowing us to make efficient use of exascale 
High Performance Computing (HPC) architecture.

The flexible nature of this block-based AMR allows us to efficiently resolve features developing 
during the BNS merger, such as small scale structures in the magnetic field instabilities, without 
requiring large scale refinement of coarser features.

To validate the performance of our code, we perform a series of tests on known physical configurations,
demonstrating the evolution of the configuration in line with expected results, in various challenging 
regimes, as well as convergence properties of the code. We also investigate the comparative performance of various choices of
methods available within \GRAthena.

In Section~\ref{sec:formalism} we introduce the equations of GRMHD in the form that \GRAthena solves,
along with key diagnostic quantities. In Section~\ref{sec:methods} we describe the key features of 
our numerical approach to solve these equations, recapping the structure of \GRAthena
and describing new additions to the code base. In Section~\ref{sec:ss} we describe the results of
 tests of \GRAthena on spacetimes with single NSs, both static and rotating, with and without 
magnetic fields, and investigate the long-term evolution of isolated magnetised stars. 
In Section~\ref{sec:collapse} we describe the results of tests on an unstable single 
star spacetime which collapses to form a BH. In Section~\ref{sec:bns} we perform BNS mergers, comparing
our results with the established \BAM code, and demonstrating the ability of \GRAthena to efficiently 
capture magnetic field amplifications using AMR. In Section~\ref{sec:scaling_tests} we show strong and weak 
scaling tests for \GRAthena for various problems, on multiple machines.

Note that in all that follows we use cgs units, with the exception of quantities derived from 
gravitational wave strains and black hole masses, which use geometric units with $G=c=\Msun = 1$,
as specified in the captions of the relevant figures.

\section{GRMHD Equations}\label{sec:formalism}

In order to evolve dynamical neutron star spacetimes, we evolve the Einstein Equations in the Z4c formulation, coupled with 
the General Relativistic Euler Equations written in conservation law form, to exploit high resolution shock capturing techniques.
In addition, magnetic fields are evolved with the Maxwell equations in the ideal MHD approximation. 
In \cite{Daszuta:2021ecf} we have already discussed the implementation of the Z4c formulation of the Einstein Equations within \GRAthena, 
and we refer the reader there for a detailed explanation, while here we will discuss the implementation of the equations describing first
General Relativistic Hydrodynamics (GRHD), and then GRMHD, within \GRAthena, and the coupling of spacetime to matter evolution.

\subsection{GRHD Equations}\label{sec:grhdeqs}

We work within the standard $(3+1)D$ decomposition of a $4D$ spacetime,
$(\mc{M},g_{\mu\nu})$~\footnote{The Lorentzian 4-metric
$g_{\mu\nu}$ has signature (-,+,+,+) and $\mu,\nu=0,1,2,3$.}
by defining a scalar field $t$
and 3 dimensional spacelike hypersurfaces $\Sigma_t$ of constant $t$
with normal vector $n^\mu$. In adapted coordinates, the induced
spatial metric is $\gamma_{ij}$ ($i,j=1,2,3$) and the extrinsic
curvature arising from their embedding in the 4D manifold is
$K_{ij}$. The lapse function and shift vector are indicated as 
$\alpha$ and  $\beta^i$ respectively.

With the spacetime, we consider a perfect fluid described by the stress energy tensor,
\begin{eqnarray}
\label{eq:Tmunu_hd}
T^{\mu\nu} = \rho h u^\mu u^\nu + p g^{\mu\nu},
\end{eqnarray}
characterised by the fluid 4-velocity, a timelike 4-vector, $u^\mu$; the fluid rest mass density $\rho$; the fluid pressure $p$; and the fluid relativistic specific enthalpy 
$h = 1 + \epsilon + \frac{p}{\rho}$, where $\epsilon$ is the specific internal energy of the fluid.
The fluid evolution equations are given by the conservation of rest mass density and the conservation
of the stress-energy tensor through the Bianchi identities,
\be
\nabla_\mu(\rho u^\mu) &=& 0,\\
\nabla_\mu T^{\mu\nu} &=& 0.
\ee
By projecting these equations onto and normal to $\Sigma_t$ we can
write these equations in conservation law form~\citep{Banyuls:1997zz}
\be 
\label{eq:claw}
\pd_t \mathbf{q} + \pd_i \mathbf{F}^i = \mathbf{s},
\ee
where the conservative variables are associated to the Eulerian
observer, moving along $n^\mu$, 
\be
\label{eq:cons_def}
\mathbf{q} = (D, S_i, \tau) = \sqrt{\gamma}\left(\rho W, \rho h W \tilde u_i, \rho h W^2 - \rho W - p \right), \nn \\
\ee
the flux vectors are,
\be
\label{eq:flux_def}
\mathbf{F}^i = 
\begin{pmatrix}
&D \alpha \tilde v^i  \\
& S_j \alpha \tilde v^i + \delta^i_j p \alpha \sqrt{\gamma}\\
& \tau \alpha \tilde v^i + \alpha \sqrt{\gamma} p v^i 
\end{pmatrix},
\ee
\begin{widetext}
and the source terms are,
\be
\label{eq:hydsrc_def}
\mathbf{s} = \alpha \sqrt{\gamma}
\begin{pmatrix}
& 0   \\
& T^{00}\left(\half \beta^i\beta^j \pd_k \gamma_{ij} - \alpha \pd_k \alpha  \right) + T^{0i} \beta^j \pd_k \gamma_{ij} + T^0_i \pd_k \beta^i + \half T^{ij}\pd_k \gamma_{ij}\\
& T^{00}\left(\beta^i\beta^j K_{ij} - \beta^i \pd_i \alpha  \right) + T^{0i} \left(2 \beta^j K_{ij} - \pd_i \alpha\right) +  T^{ij}K_{ij}
\end{pmatrix}.\nn \\
\ee
\end{widetext}
Here we have defined $\gamma = \mathrm{det}(\gamma_{ij})$, and the
Lorentz factor between Eulerian and comoving fluid observers $W =
-n_\mu u^\mu$. Further, we can write the fluid 4-velocity as $u^\mu =
(u^0, u^i)$, with the spatial components as projected onto $\Sigma_t$
defined as, 
\be
\tilde u^i = \bot^i_\mu u^\mu = u^i + \frac{W \beta^i}{\alpha}\,,
\ee
with the projection operator $\bot^i_{\mu} = \delta^i_\mu + n^i n_\mu$.
Following common notation in the literature we also define the fluid
3-velocity $v^i =  \frac{\tilde u^i}{W}$ and $\tilde v^i = v^i -
\frac{\beta^i}{\alpha}$. 
The fluxes, Eq.~\eqref{eq:flux_def}, depend on a set
of primitive variables defined in the comoving fluid frame. Here we
take as our fundamental primitive variables  
$\mathbf{w} = (\rho, \tilde u^i, p)$, and derive other fluid variables
from this set. 

We note that in \Athena the spacetime metric is assumed to be
stationary, and so the implementation of these equations in
conservation law form can be simplified.
First, the densitisation of the conservative variables by
$\sqrt{\gamma}$ can be removed since the determinant can be pulled out of the time derivative 
in Eq.~\eqref{eq:claw}
and multiplied as a time independent factor after the time
integration. Second, by choosing the conservative variables as the
components of the stress-energy tensor $T_0^\mu$, the source terms for
$T_0^0$ become only time derivatives of the metric, which vanish. In
order to evolve GRMHD problems in dynamical spacetimes we switch
to the more general formulation presented above. 
 
Since the stress-energy tensor Eq.~\eqref{eq:Tmunu_hd} enters the
right hand side of the Einstein equations, the projections
of this tensor enter the right hand
sides of the Z4c equations, specifically Eqs~(10-13)
of~\cite{Daszuta:2021ecf}. 
For the perfect fluid stress-energy tensor these are given by,
\begin{subequations}
\be
\rho_{\mathrm{ADM}} &=&  \rho h W^2 - p,\\
S_{i~\mathrm{ADM}} &=&  \rho h W^2 v_i,\\
S_{ij~\mathrm{ADM}} &=&  \rho h W^2 v_i v_j + p \gamma_{ij}\,.
\ee
\end{subequations}
Note the addition of the subscript ADM here in contrast to
\cite{Daszuta:2021ecf} to avoid confusion with other variables used
in this paper.

\subsection{GRMHD Equations}

In order to extend the above system to GRMHD, 
we introduce the antisymmetric electromagnetic tensor $F_{\mu\nu}$,
which can be decomposed with respect to an Eulerian observer's
4-velocity $n^\mu$ as follows, 
\be
F^{\mu\nu} = n^\mu \mc{E}^\nu - \mc{E}^\mu n^\nu -  \eps^{\mu\nu\rho\sigma} n_\rho \mc{B}_\sigma\,,
\ee
where $\mc{E}^\mu, \mc{B}^\mu$ are the electric and magnetic field measured by the
Eulerian observers, and $\eps^{\mu\nu\rho\sigma}$ is the 4 dimensional
Levi-Civita tensor.
The electromagnetic tensor can be similarly decomposed along the
fluid 4-velocity $u^\mu$, in order to obtain the magnetic field
components for a comoving observer, given by the 
4-vector $b^\mu = u_\nu{}^*F^{\mu\nu}$, with the dual tensor ${}^*F_{\mu\nu} = \frac{1}{2}\eps^{\mu\nu\rho\sigma}F_{\rho\sigma}$.
The relationship between the two representations of the magnetic field
is,
\begin{subequations}
\be
b^0 &=& \frac{W \mc{B}^i v_i}{\alpha}, \\
b^i &=&  \frac{\mc{B}^i + \alpha b^0 W \tilde v^i}{W},\\
b^2 &=& g_{\mu\nu} b^\mu b^\nu = \frac{(\alpha b^0)^2 + \mc{B}^i \mc{B}^j \gamma_{ij}}{W^2}\,.
\ee
\end{subequations}
The full stress-energy tensor for the fluid and electromagnetic fields is,
\be
\label{eq:EMT_MHD}
T^{\mu\nu} = (\rho h + b^2) u^\mu u^\nu + \left(p + \frac{b^2}{2}\right) g^{\mu\nu} - b^\mu b^\nu\,.
\ee

We note that a factor of $\sqrt{4\pi}$ has been absorbed into the magnetic field definition.

We work in the limit of ideal MHD, where
resistivity is zero, and so consequently in the comoving fluid frame
the electric field vanishes, \ie~$u^\mu F_{\mu\nu} = 0$. 
Under this assumption Maxwell's equations are 
\be
\label{eq:Maxwell}
\nabla_\mu {}^*F^{\mu\nu} = 0.
\ee

By projecting Eq.~\eqref{eq:Maxwell} into $\Sigma_t$ we obtain
Maxwell's equations in conservation law form and the full GRMHD system.
The primitive variables become $\mathbf{w} = (\rho, \tilde u^i, p, B^i)$, and the conservatives,
\be
\label{eq:consB_def}
\mathbf{q} &=& (D, S_j, \tau, B^k)\\ &=& \sqrt{\gamma}(\rho W, (\rho h
+ b^2) W \tilde u_j - \alpha b^0 b_j,\\&& (\rho h + b^2) W^2 - \rho W
- \left(p + \frac{b^2}{2}\right) - (\alpha b^0)^2, \mc{B}^k  )\,. \nn 
\ee
The flux vector becomes
\be
\label{eq:fluxB_def}
\mathbf{F}^i = 
\begin{pmatrix}
&D \alpha \tilde v^i  \\
& S_j \alpha \tilde v^i + \delta^i_j \left(p + \frac{b^2}{2}\right) \alpha \sqrt{\gamma} - \frac{\alpha \sqrt{\gamma} b_j \mc{B}^i}{W}\\
& \tau \alpha \tilde v^i + \alpha \sqrt{\gamma} \left(\left(p + \frac{b^2}{2}\right) v^i  - \frac{\alpha b^0 \mc{B}^i}{W}\right) \\
& \alpha (B^k \tilde v^i - B^i \tilde v^k)
\end{pmatrix}.
\ee
We note that the expression for the sources Eq.~\eqref{eq:hydsrc_def}
remains valid, with the stress energy tensor now given by
Eq.~\eqref{eq:EMT_MHD} and the sources for the magnetic field components
zero. 
The ADM sources are,
\begin{subequations}
  \be
\rho_{\mathrm{ADM}} &=&  (\rho h + b^2) W^2 - \left(p + \frac{b^2}{2}\right) - (\alpha b^0)^2, \\
S_{i~\mathrm{ADM}} &=&  (\rho h + b^2) W^2 v_i - b^0 b_i \alpha,\\
S_{ij~\mathrm{ADM}} &=&  (\rho h + b^2) W^2 v_i v_j + \left(p+ \frac{b^2}{2}\right) \gamma_{ij} - b_i b_j\,. \nn\\
\ee
\end{subequations}
The final additional equation to consider is the elliptic equation obtained by projecting Eq.~\eqref{eq:Maxwell} normal to $\Sigma_t$, 
\be
\label{eq:gausslaw}
\pd_i B^i = 0\,,
\ee
\ie~the divergence-free constraint on $B^i$. This will be
automatically enforced by the discretisation of the equations as
detailed in Section~\ref{sec:CT}. 

We note that for the GRMHD system we calculate the sound speeds following 
\cite{Gammie:2003rj} with an approximated quadratic dispersion relation.

\subsubsection{EOS}

The above system of conservation laws determines the evolution of 5
(7) degrees of freedom for GR(M)HD. However, the fluid is determined
by 6 (8) primitive variables $(\rho,p,\epsilon,\tilde u^i, B^i)$
(with one degree of freedom in the magnetic field constrained by Eq.~\eqref{eq:gausslaw}) 
. The system is closed by an
equation of state (EOS) that establishes the thermodynamical
relationship between $(\rho, p, \epsilon)$. In this study, we 
restrict our attention to the cases of an ideal gas and a
3-dimensional tabulated equation of state. 
For the ideal gas the
pressure is described by the Gamma law EOS,
\begin{eqnarray}
p = \rho\epsilon (\Gamma - 1),
\end{eqnarray}
where $\Gamma$ is related to the adiabatic index $n$ of the fluid
through the relation $\Gamma = 1 - 1/n$. 
In order to initialise the pressure we also use a
barotropic equation of state, 
\begin{eqnarray}
p = K\rho^\Gamma,
\end{eqnarray}
with the dimensionful parameter $K$ introducing a mass scale for the
fluid. In this paper $\Gamma = 2$ always, while $K$ will vary from
problem to problem. 

For finite-temperature tabulated EOSs we use the temperature $T$
instead of $\epsilon$ as the ``thermal'' primitive variable so as to make
better contact with nuclear physics calculations.
We also introduce a number of species fractions $Y_i$ to track the
composition of the fluid, following the CMA scheme of \cite{Plewa:1998nma}. In this work, we limit these species fractions to only the
electron fraction $Y_\mathrm{e}$, which, for a  
charge neutral fluid composed of neutrons, protons, and electrons, is given by 
\be
Y_\mathrm{e} = \frac{n_\mathrm{e}}{n_\mathrm{b}} = \frac{n_\mathrm{p}}{n_\mathrm{b}} = Y_\mathrm{q}\,,
\ee
where $n_\mathrm{e}$, $n_\mathrm{b}$, and $n_\mathrm{p}$ are the
electron, baryon, and proton number densities respectively, and
$Y_\mathrm{q}$ is the charge fraction. With this extra degree of
freedom ($Y_\mathrm{e}$) we require an extra evolved variable in  
the GRMHD equations, namely $DY_\mathrm{e} = D \cdot
Y_\mathrm{e}$. The flux term for this extra variable is
obtained by multiplying the corresponding equations for the conserved
density by the electron fraction, and the source term remains zero. 
The pressure relationship for this equation of state becomes
\be
p = p\left( \rho, T, Y_\mathrm{e} \right)\,,
\ee
where the right hand side is calculated by 3-dimensional linear interpolation
of $\log p$ tabulated in $(\log \rho, \log T, Y_\mathrm{e})$. In
addition to the pressure, other thermodynamic variables required
throughout the simulation (namely energy density $e$, defined through $\log e = \log(\rho(1 + \epsilon))$ and
the squared sound speed $c^2_s$) are also tabulated in this
manner. 
Our EOS implementation supports tables in the form used by the CompOSE database \citep{Typel:2013rza}
converted to the 
HDF5 format and modified to include the sound speed by the PyCompOSE tool \footnote{https://bitbucket.org/dradice/pycompose/}. 

\subsection{Diagnostic quantities}

We list here common diagnostic quantities referred to in later sections.
The baryon mass is defined as 
\be\label{eq:Mb}
M_b = \int_{\Sigma_t} D \d^3 x\,,
\ee
and it is a conserved quantity. The integral is calculated on the
entire computational domain.

We measure various energy contributions to the GRMHD system as follows
\begin{subequations}
\begin{eqnarray}
E_{\rm Kin} &=&\frac{1}{2} \int_{\Sigma_t} \frac{S_iS^i}{D} \d^3 x,\\
E_B &=&\frac{1}{2}\int_{\Sigma_t} \sqrt{\gamma}W b^2\d^3 x, \\
E_{\rm int} &=&  \int_{\Sigma_t} D\epsilon \d^3 x\,.
\end{eqnarray}
\end{subequations}

The gravitational wave strain is decomposed in multipoles
$h_{\ell m}(t)$ as
\be
h_+ - ih_\times = \frac{G}{D_L} \sum_{\ell\geq 2}\sum_{m=-\ell}^\ell
h_{\ell  m}(t)\,{}_{-2}Y_{\ell m}(\theta, \phi)\,,
\ee
where ${}_{-2}Y_{\ell m}(\theta, \phi)$ are the spin-weighted spherical
harmonics on the sphere following the convention of \cite{Goldberg:1966uu}
up to a Condon-Shortley phase factor of $(-1)^m$
, and $D_L=(1+z)R$ is the luminosity distance
of a source located at distance $R$ to the observer and at redshift $z$.
The GW are extracted on coordinate spheres centered at the grid
origin sampled using geodesic spheres constructed with $n_Q = 10$ levels of refinement, see \cite{Daszuta:2021ecf} for details.

We define also the convergence factor,
\begin{eqnarray}\label{eq:convfactor}
Q_n = \frac{\delta x_c^n - \delta x_m^n}{\delta x_c^n - \delta x_f^n}
\end{eqnarray}
used to demonstrate self convergence for a set of coarse, medium and fine runs with
finest grid spacings $(\delta x_c,\delta x_m,\delta x_f)$ respectively at order $n$.

\section{Methods}\label{sec:methods}

\subsection{Mesh}
\Athena uses a block-based oct-tree AMR to construct its computational domain,
known as the \Mesh. The structure of this \Mesh{} is detailed in \cite{Stone:2020,Daszuta:2021ecf},
and so here we simply recall the key parameters controlling the structure of the \Mesh{}
which will be referred to throughout the paper. Before any mesh refinement is performed,
the \Mesh{} of predetermined coordinate extent is sampled by $[N^x,N^y,N^z]$ cells. 
This \Mesh{} is then subdivided into \MeshBlock s which consist of $[N^x_B,N^y_B,N^z_B]$ 
cells, where $N^i_B | N^i$. For most of the applications below we select $N^x=N^y=N^z:=N$ and $N^x_B=N^y_B=N^z_B:=N_B=16$.
Note that the number of vertices in a \MeshBlock{} is always one larger than the number of cells.
When mesh refinement occurs, a \MeshBlock{} is replaced with (in 3D) 8 child \MeshBlock s, with the
same number of cells as the parent block, but half the coordinate extent in each direction,
thus doubling the spatial resolution. In this way \MeshBlock s can be refined multiple times
to achieve a target resolution, with the only restriction that neighbouring \MeshBlock s may only 
differ by at most 1 level of refinement. Finally, we emphasise that $N$ is the parameter that controls 
the base resolution of the unrefined grid, and so is the parameter we use below during tests 
to characterise resolution in convergent series.

Each grid cell exists on only one \MeshBlock, with the exception of shared outer vertices,
with no Berger-Oliger time subcycling employed for evolutions with mesh refinement. We fix the CFL
factor to 0.25, setting a global timestep $\delta t$ as determined by the grid spacing $\delta x$ on
the finest \MeshBlock{}.

\subsection{Intergrid communication} \label{sec:intergrid}

As detailed in \cite{Daszuta:2021ecf}, \GRAthena uses a vertex-centred (VC) 
grid upon which the Z4c variables are discretised, in order to allow for
efficient performance of restriction and prolongation operations between
neighbouring refinement levels when using high-order operators on a refined
mesh. In contrast \Athena uses a cell centred (CC) grid to discretise the volume 
averaged conserved quantities of the Euler equations, in order to employ 
standard conservative Godunov based Finite Volume methods used in order to 
capture shocks forming in the fluid. 
Further, \Athena uses a face centred (FC) and edge centred (EC) 
discretisation of the magnetic and electric fields respectively, 
in order to use the Constrained Transport 
algorithm in the evolution of the magnetic field equations while preserving Eq.~\eqref{eq:gausslaw}.
Since the MHD equations require the value of metric variables to perform densitisations, 
and the Z4c equations require MHD variables to calculate the sources in the 
stress-energy tensor, we interpolate these variables between these representations
using Lagrangian interpolation, the order of which we are able to vary between 
variables.

\subsection{Constrained Transport scheme}
\label{sec:CT}
In order to evolve the magnetic field while preserving the divergence-free constraint
in Eq.~\eqref{eq:gausslaw}, \Athena uses the Constrained Transport algorithm initially developed by
\cite{Evans:1988a}, and further developed and utilised in 
\cite{Gardiner:2005,Gardiner:2007nc,Stone:2008mh,Beckwith:2011iy} 
. The specific implementation 
considered follows Section 3.5 of \cite{White:2015omx} 
, as included in \Athena for stationary 
spacetimes.  The implementation in \GRAthena matches this, with the exception of volume elements 
appropriately included in variable definitions as opposed to face area or edge length weightings, 
as discussed in Section~\ref{sec:formalism}.

To ensure preservation of this constraint through simulations with AMR, where magnetic field
quantities must be interpolated to new \MeshBlock s as they are dynamically created, \Athena
uses the curl and divergence preserving restriction and prolongation operators of \cite{Toth:2002a}
and implements corrections to fluxes and electric fields as detailed in Section (2.1.4-2.1.5) of \cite{Stone:2020}.

\subsection{Reconstruction}

In order to evaluate the numerical fluxes at the cell interfaces, we reconstruct
the primitive variables
from their cell centred to a face centred representation.
In order to perform this reconstruction 
while avoiding introducing an increase in total variation, we employ a variety of limited schemes.
Within this paper we will present results obtained using the Piecewise Linear Method (PLM) \citep{vanLeer:1974a}, and Piecewise Parabolic 
Method (PPM) \citep{Colella:2011a,McCorquodale:2015a}, as implemented in \Athena \citep{Stone:2020}, as well as 2 implementations of the Weighted Essentially Non-Oscillatory 
(WENO) schemes \citep{Liu:1994}, WENO5 \citep{Jiang:1996a} and WENOZ \citep{Borges:2008a}, following the implementation in the \BAM code \citep{Thierfelder:2011yi}.

\subsection{Conservative-To-Primitive Variable Inversion}

In order to evaluate the fluxes in Eq.~\eqref{eq:claw} we convert the
conservative variables, $\mathbf{q}$ to their primitive representation
$\mathbf{w}$ by inverting the definitions of $\mathbf{q}$ in
Eq.~\eqref{eq:cons_def}.  
This results in a non-linear system of equations which must be solved 
numerically for the primitive variables, for which there are a number
of commonly employed strategies, see \cite{Siegel:2017sav} for a 
summary and comparison of a selection of approaches.

To perform this inversion, \GRAthena implements both the external library
\reprimand  \citep{Kastaun:2020uxr} and a similar custom implementation
called \PrimitiveSolver described in Appendix~\ref{app:PrimitiveSolver}.
The conservative-to-primitive inversion operates on a pointwise basis and so
operates independently of the structure of the \Mesh\, of \GRAthena.
In our simulations we employ a tolerance of $10^{-10}$ in the
bracketing algorithm of \reprimand, and find that robust performance
requires a ceiling of 20 on the velocity variable $\tilde
u$. Points for which the conserved to primitive variable inversion
fails, or at which the ceiling values of variables are exceeded are
set to atmosphere values (see below).

Physically we expect a NS to have a well defined, sharp, surface, over
which the fluid density drops to zero. Resolving such a sharp feature can introduce
numerical instabilities, and the presence of a region with zero fluid density 
will cause weak solutions to the Euler equations to become ill-defined.
To avoid these issues we follow the standard technique of introducing
a low-density fluid atmosphere which fills the entire computational
domain outside of the NS.
The primitive variables within the atmosphere are set by,
\begin{subequations}
\be
\rho_{\mathrm{atm}} &=& f_{\mathrm{atm}} \rho_{\mathrm{max}}, \\
p_{\mathrm{atm}} &=& p(\rho_{\mathrm{atm}}),\\
\tilde u^i_{\mathrm{atm}} &=& 0\,,
\ee
\end{subequations}
with $\rho_{\mathrm{max}}$ the maximum value of the fluid density in
the initial data, and $f_{\mathrm{atm}}$ a parameter chosen for the
specific problem. We do not alter the value of the magnetic field in
atmosphere.  
As the initial data evolves, the fluid profile of the NS will begin to
spread out due either to numerical dissipative effects, or due to physical 
disruption of the star. Here, the fluid can flow into atmosphere
regions, and it is necessary to define a criterion to tag given
numerical cells as atmosphere. We set a cell to atmosphere if the
fluid density falls below the value
\be
\rho_{\mathrm{thr}} = f_{\mathrm{thr}} \rho_{\mathrm{atm}},
\ee
where again $f_{\mathrm{thr}}$ is a free parameter. We also set
atmosphere values in the case that the conservative-to-primitive
variable conversion fails to converge to a solution, or when the upper
bound on the velocity is reached. When atmosphere values are set, we recalculate
the conservative variables from the atmosphere primitives and metric variables.
Within this paper we set $f_{\mathrm{thr}} = 1$ always, while
$f_{\mathrm{atm}}$ varies between $10^{-13} - 10^{-18}$.

\section{Single Neutron Star spacetimes}\label{sec:ss}

This section presents tests that involve the evolution of 
stable NS spacetimes. We first discuss the evolution of static and
rotating NSs, the convergence of various diagnostics and the
performances of reconstruction schemes in maintaining the initial
equilibria configurations. We then present long-term simulations of
magnetic field instabilities extending our previous work in the Cowling
approximation in \cite{Sur:2021awe}.

We note that in all the tests presented below in Sections~\ref{sec:ss} - \ref{sec:scaling_tests} we use a 3rd order 
Runge-Kutta time integrator, 6th order accurate finite differencing
operators in the right-hand side of the Einstein equations, set
the damping parameters in the Z4c equations 
$\kappa_1 = 0.02, \kappa_2 = 0$, and set the magnitude of Kreiss-Oliger
dissipation with the parameter $\sigma_{KO} = 0.5$, parameters which are defined in 
\cite{Daszuta:2021ecf}.

\subsection{GRHD evolution of Static Neutron Star }\label{sec:ss:sns_grhd}

We start with the evolution of initial data made of a static NS of
baryon mass $M_b=1.506\Mo$ and gravitational mass $M=1.4\Mo$. The
initial data is computed by solving the Tolman-Oppenheimer-Volkoff
(TOV) equations using a central rest-mass density of
$\rho_c=7.905\times10^{14}\gccm$, using an ideal gas EOS with $\Gamma=2$, with the
initial pressure set with the polytropic EOS with $K=100$. This NS model is called A0 and has
been used in several previous code tests, \eg~\citep{Font:2001ew,Thierfelder:2011yi}.

The outer boundaries of \GRAthena{} grid are placed at $[\pm 378.0,\pm
  378.0,\pm 378.0]$~km, with 6 levels of static mesh refinement centred on
the star. The innermost level has extension $[\pm 14.8, \pm 14.8, \pm 14.8]$~km
and covers the NS entirely. 
We consider simulations at resolutions $N=32,48,64,128$, corresponding to a grid spacing
on the finest level of $[369,246,185,92.3]$~m and various
reconstruction schemes.

 \begin{figure}[t]
   \centering 
     \includegraphics[width=0.49\textwidth]{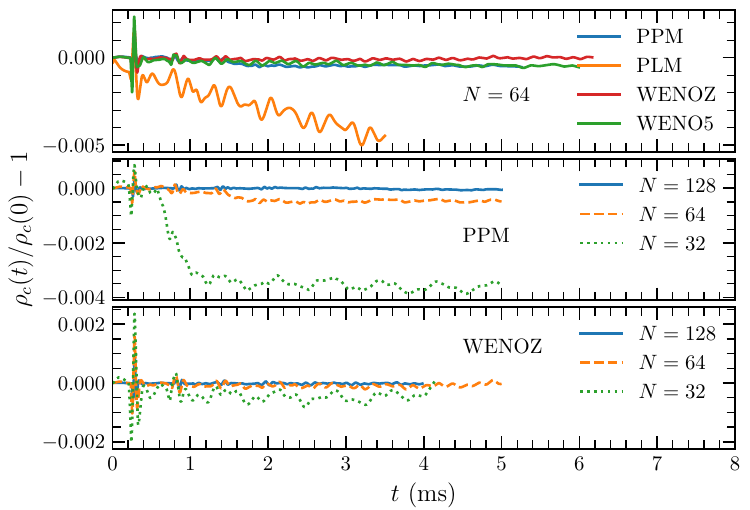}
     \caption{GRHD Evolution of central rest-mass density of the static
       star A0 model. Top: varying reconstruction at constant resolution
       $N=64$. Central: varying resolution for PPM reconstruction.
       Bottom: varying resolution for WENOZ reconstruction. Some data
       is cut to $5$~ms for visualization purposes.
     }
     \label{fig:A0_rhoc}
 \end{figure}

Figure~\ref{fig:A0_rhoc} shows the evolution of the central rest-mass
density $\rho_c(t)$ for various reconstructions schemes and (for PPM
and WENOZ reconstructions) for various resolutions. All the evolutions
are stable over a period of several crossing times. The central
rest-mass density shows the characteristic oscillations triggered by
truncation errors. The latter are larger at the star surface, where a
non-zero velocity field develops quickly. The oscillation frequencies
extracted from the power spectra are compatible with the first three radial
modes, namely $\nu_F=1462,3938$ and $5928$~Hz as computed in linear
perturbation theory~\citep{Baiotti:2008nf}.
The oscillations' amplitude is below the percent level and converge to
zero with increasing resolution. Low resolutions typically result in
an unphysical expansion of the star beyond its initial radius. The use
of the PLM reconstruction produces the largest expansion of the star,
and it is effectively the least accurate scheme as noted also
elsewhere, \eg~\citep{Thierfelder:2011yi}. 
The PPM scheme shows also a prominent drift in $\rho_c(t)$ at low 
resolutions ($N=32$) but for higher resolutions it retains a similar
accuracy to the WENO schemes. 

 \begin{figure}[t]
   \centering 
     \includegraphics[width=0.49\textwidth]{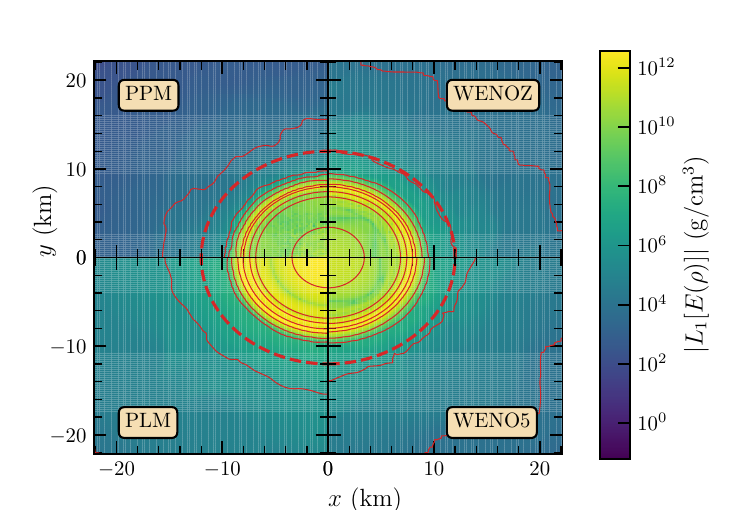}
     \caption{$L_1$ norm on the $x-y$ plane of the rest-mass density
       error of the static star model A0 for varying reconstruction. The
       dashed red line corresponds to the initial NS radius,
       contour levels corresponds to rest-mass densities of
       $\rho=[10^4,10^6, 10^8, 10^{10}, 10^{12}, 10^{13}, 5\times 10^{13}, 10^{14}, 5\times10^{14}]~\gccm$. 
       Data refer to GRHD evolutions at time $t=3.45$~ms and at at
       constant resolution $N=64$.}
  \label{fig:A0_grhdrecon2d}
 \end{figure}

In Figure~\ref{fig:A0_grhdrecon2d} we visualise the $L_1$ normed
error of $\rho$, 
\be
 L_1[E(\rho)](\mathbf{x},t) :=  |\rho(\mathbf{x},t) - \rho(\mathbf{x},0)|,
\ee
as a two-dimensional slice for the 4 different reconstruction
schemes and $N=64$. This plot shows complementary information to
the central density oscillation. The error of the PLM scheme is at all
times worse than that in the other three schemes. The PPM scheme
performs the worst at preserving the sharp star surface at early times
(not shown in the plot), but at later times it has an overall lower
error both at the surface and within the star itself away from the centre
than the WENO schemes. 

 \begin{figure}[t]
   \centering 
     \includegraphics[width=0.49\textwidth]{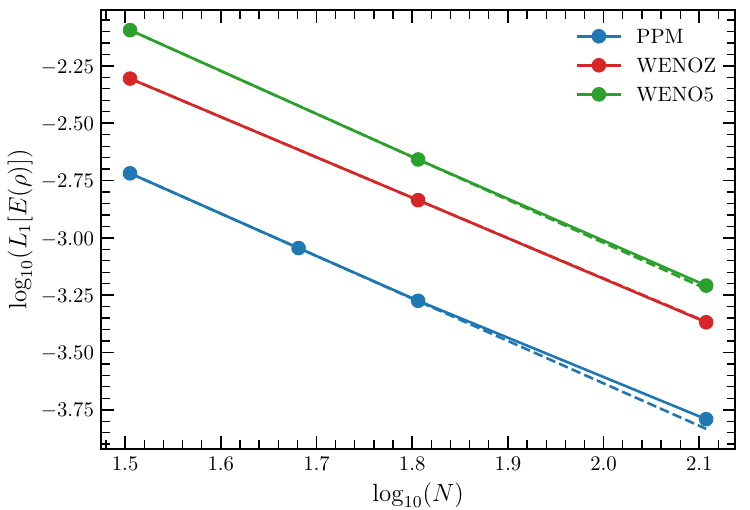}
     \caption{Convergence of $L_1[E(\rho)]$ norm of static star model A0 
       at $t=2.96$ ms.
       Data refer to GRHD evolutions and various reconstruction
       schemes. The modulus of the slopes, corresponding to the order of
       convergence is $1.85,1.76,1.87$ for PPM, WENOZ, WENO5 respectively.
     }
  \label{fig:A0_L1rho}
 \end{figure}

The convergence of some of these results in the 
$L_1[E(\rho)]$ norm integrated on the whole computational domain are
shown in Figure~\ref{fig:A0_L1rho}, with dashed lines showing linear 
extrapolations from the first two data points, the slopes of which
indicate the order of convergence. 
All errors in the 
simulations converge at approximately second order rate. Consistent
with the above discussion, the PPM scheme has the lowest absolute
errors at the considered resolutions, with the convergent behaviour most 
consistent at all resolutions in the WENO simulations, especially WENOZ.
At the lowest resolutions the WENO results stay within the convergent regime 
for the  full duration of the simulation, whereas this does not happen for the 
lowest resolution PPM simulation, as seen in the lower two panels of Figure~\ref{fig:A0_rhoc}. 

 \begin{figure}[t]
   \centering 
     \includegraphics[width=0.49\textwidth]{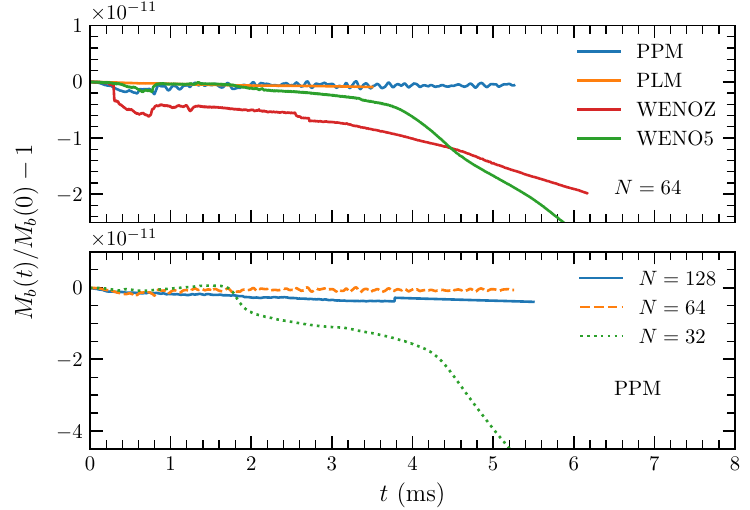}
     \caption{Baryon mass conservation along the GRHD evolution of static
       star model A0. The mass conservation can be violated due to the
       artificial low-density atmosphere as the fluid expands beyond
       the computational domain.}
  \label{fig:A0_masscons}
 \end{figure}

Figure~\ref{fig:A0_masscons} reports the relative violation of the
baryon mass conservation within the simulation. The mass is conserved
at the $10^{-11}$ relative level for all the schemes at the considered
resolutions. The mass conservation is slightly violated due to the
artificial low-density atmosphere and as the fluid expands beyond the
computational domain. Hence, higher resolutions can reduce the
violation by better resolving the surface and reducing the star's
expansion. In this respect, the PPM scheme performs best among those
explored. 
 
 \begin{figure}[t]
   \centering 
     \includegraphics[width=0.49\textwidth]{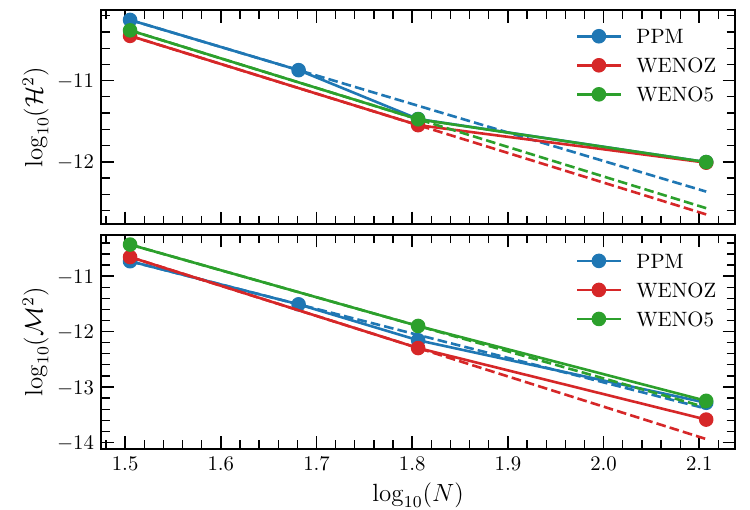}
     \caption{Top: Convergence of the $L_1$ norm of the Hamiltonian constraint of static star model A0.
       Bottom: Convergence of the $L_1$ norm of the momentum constraint of static star model A0.
       Data is evaluated at $t=2.96$ ms. The 
       modulus of the slopes, corresponding to the order of convergence 
       in the Hamiltonian (momentum) constraint is 
       $3.51 (4.41), 3.65 (5.46), 3.64 (4.88)$ for PPM, WENOZ, WENO5 respectively.} 
     \label{fig:grhdconvh}
 \end{figure}

Finally, we inspect the convergence of the constraints in Einstein's
equations (see Eqs 18-20 in \cite{Daszuta:2021ecf}).
The key property of the Z4c system is to propagate and damp the
constraints during the evolution
\citep{Bernuzzi:2009ex,Hilditch:2012fp}. For example, the norm of the
Hamiltonian constraint is of order ${\sim}10^{-11}$
(${\sim}10^{-14}$) for the lowest $N=32$ (highest $N=128$)
resolutions by end of the simulations. This value is typically several
orders of magnitude smaller than the constraint violation in the
initial data. For 
this reason, convergence properties might be difficult to interpret.
Figure~\ref{fig:grhdconvh} shows the $L_1$-norm convergence of the
Hamiltonian and momentum constraints by the end of the simulations. 
The observed convergence is approximately 4th or 5th order, depending on the reconstruction method,
 indicating that the finite differencing scheme used in the evaluation of the RHS of the
Z4c equations is a significant source of error at this time. 

 \subsection{GRHD evolution of Rotating Neutron Star}\label{sec:ss:rns_grhd}

Next, we consider the evolution of a stable uniformly rotating equilibrium
configuration close to the mass shedding limit. Specifically, we evolve the
model AU4 of \cite{Dimmelmeier:2005zk} that is rotating at a frequency of
655~Hz, with a corresponding rotational period of 1.53 ~ ms. The initial data has baryon mass $M_b  = 1.506\Mo$ and 
gravitational mass $M=1.415\Mo$, and the same EOS as in Section~\ref{sec:ss:sns_grhd}.
Initial data are generated using the \texttt{RNS} code of
\cite{Stergioulas:1994ea}, which is incorporated into \GRAthena{}
as an external shared library \footnote{The code can be found at
\url{https://bitbucket.org/bernuzzi/rnsc/src/master/}.}.
The model is characterised by a central density of $\rho_c = 5.453\times10^{14}\gccm$
and a polar-to-equatorial coordinate axis ratio of $r_p/r_e =0.698$.
This is a challenging test where all the metric-matter source terms are
non-zero and involving a sharp velocity profile at the star surface.

 \begin{figure}[t]
   \centering 
     \includegraphics[width=0.49\textwidth]{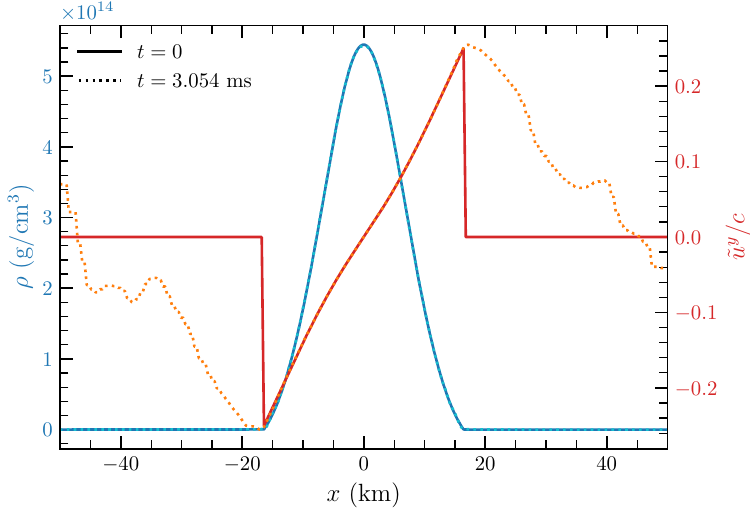}
     \caption{Profiles of velocity component $\tilde u^y$ and rest-mass
       density $\rho$ along the $x$ direction at the initial time and
       after two rotational periods. Data refer to resolution
       $N=32$ with PPM reconstruction. Note that velocity is plotted in geometric units. 
     }
  \label{fig:AU4_profiles}
 \end{figure}

Here we use the same grid set up as in Section~\ref{sec:ss:sns_grhd} 
with the highest refinement level fully covering the star.
The hydrodynamics evolution uses the PPM and WENOZ reconstructions. We do not
consider magnetic fields in this test.

Figure~\ref{fig:AU4_profiles} shows one-dimensional profiles of the
rest-mass density and the velocity component $\tilde u^y$ for the lowest
resolution $N=32$. For each quantity we show two profiles: the
initial data (representing also the exact solution at each period) and
the evolved data after the second rotational period.
The density profile is well maintained within our evolution, 
and is visually indistinguishable from the initial profile. 
The velocity profile is
well-simulated up to the lowest density regions, while the sharp
spikes at the surface are not captured. Since no cutoff above the
level of the atmosphere is used ($f_{\rm thr} = 1$), fluid elements near the surface
expand beyond the initial star radius. This is a common deficit of
Cartesian finite-volume codes that can be improved with more
sophisticated flux and reconstruction schemes \citep{Kastaun:2006ik,Guercilena:2016fdl,Doulis:2022vkx}.
Note that however, the overall angular momentum profile is largely
unaffected by this inaccuracy. 

 \begin{figure}[t]
   \centering 
     \includegraphics[width=0.49\textwidth]{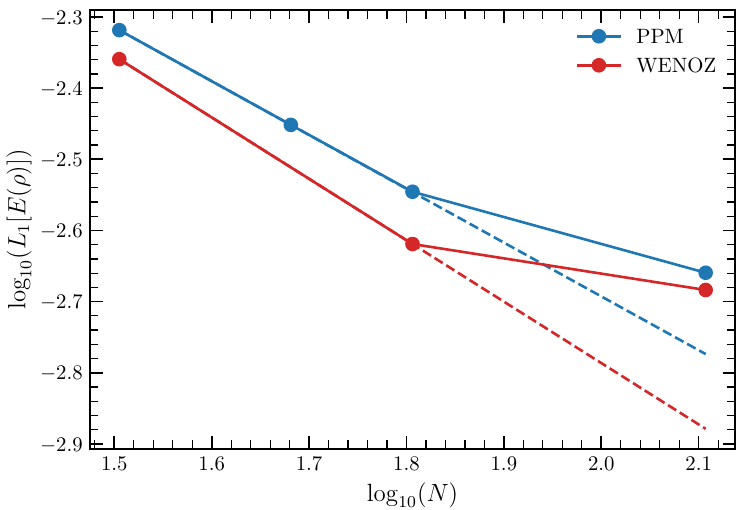}
     \caption{Convergence of $L_1$ norm of the error of the rest-mass density of
       rotating star AU4. Data refer to an
       evolution time of $t=3.06$~ms (second rotational
       period). The
       modulus of the slopes, corresponding to the order of convergence
       is $0.75, 0.86 $ for PPM, WENOZ, respectively. 
     } 
  \label{fig:AU4_L1conv}
 \end{figure}

Figure~\ref{fig:AU4_L1conv} shows the convergence of the $L_1$ norm of
the difference between the rest-mass density profile and the initial
density profile. The convergence rate is approximately first order, as
expected since the surface dynamics dominate the error source. Mass
conservation is in-line with the static star tests; relative
variations are of order ${\sim}10^{-11}$ during the entire
evolution for the lowest resolution.

\subsection{Finite-temperature EOS evolution of Static Neutron Star}\label{sec:ss:sns_grhmd_teos}

We now consider the SFHo EOS of \cite{Steiner:2012rk} from the CompOSE database \citep{Typel:2013rza} to obtain a 
configuration similar to the A0 model described above. Taking a 
constant initial temperature of $T=0.1\ \mathrm{MeV}$ and assuming 
cold $\beta$-equilibrium 
($\mu_\mathrm{n}=\mu_\mathrm{p} + \mu_\mathrm{e}$) gives us a 
1-dimensional slice of the full table, which we use with a central 
density of $\rho_c=8.523\times10^{14}\ \gccm$ to solve the TOV equations. 
The result is a star with baryon mass $M_b=1.555\ \Mo$ and 
gravitational mass $M=1.4\ \Mo$. For the evolution we use the same grid 
setup as that described in Section~\ref{sec:ss:sns_grhd}, the PPM 
reconstruction method, and the \PrimitiveSolver 
conservative-to-primitive library.

As with the baryon mass in Figure~\ref{fig:A0_masscons} 
we want to know 
to what degree the conservation of the scalar is violated. 
Analogously to Eq.~\eqref{eq:Mb}, we define the ``proton mass'' $M_p$,

\be\label{eq:Mp}
M_p = \int_{\Sigma_t} DY_{\mathrm{e}} \d^3 x\,.
\ee

Note that this is conserved only because there are no reactions present 
in the simulation, and we use the name ``proton mass'' only as a 
parallel to $M_b$. As we expect this to be coupled strongly to violations 
in the conservation of baryon mass, we take the relative error of each 
quantity, $\mathrm{err}(Q,t) = (Q(t) / Q(0)) - 1$ for $Q = M_b, M_p$, and then 
plot the ratio of these errors.

\begin{figure}[t]
  \centering 
    \includegraphics[width=0.49\textwidth]{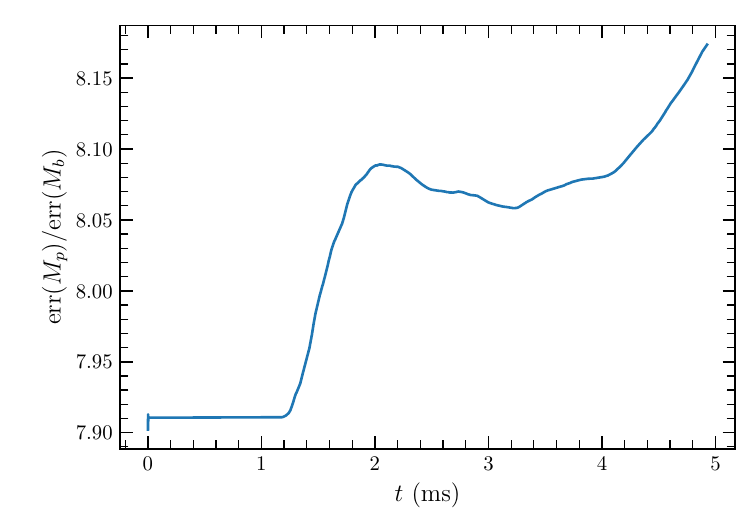}
    \caption{Ratio of the relative error in the baryon mass, $M_b$ 
             (as defined in Eq.~\ref{eq:Mb}), to the ``proton mass'', 
             $M_p$ (Eq.~\ref{eq:Mp}), for $\sim 5\ \rm{ms}$ of evolution of a 
             $M=1.4\ \Mo$ TOV star using the SFHo \citep{Steiner:2012rk} EOS.}
 \label{fig:tabul_cons}
\end{figure}

Figure~\ref{fig:tabul_cons} shows the ratio of these errors for the 
aforementioned simulation setup. We see that the ratio of errors varies 
by $\sim 20\%$ through the course of the simulation, therefore the 
conservation of the scalar is maintained to approximately the same 
degree as the conservation of mass. If the relative error in the scalar
came entirely from the error in the mass we would expect this error 
ratio to be $1$, whereas if there were some constant error of the same 
magnitude in $M_b$ and $M_p$ we would expect the ratio to be around 
$1/Y_\mathrm{e}$, which is $\sim 2$ in the atmosphere (most of the 
domain by volume), and $\sim 20$ in the core (where most of the mass 
lies) 
The rest-mass averaged value would be $\sim M_b/M_p \approx 17$. 
We see that the error falls within the range of expected values.

\begin{figure}[t]
  \centering 
    \includegraphics[width=0.49\textwidth]{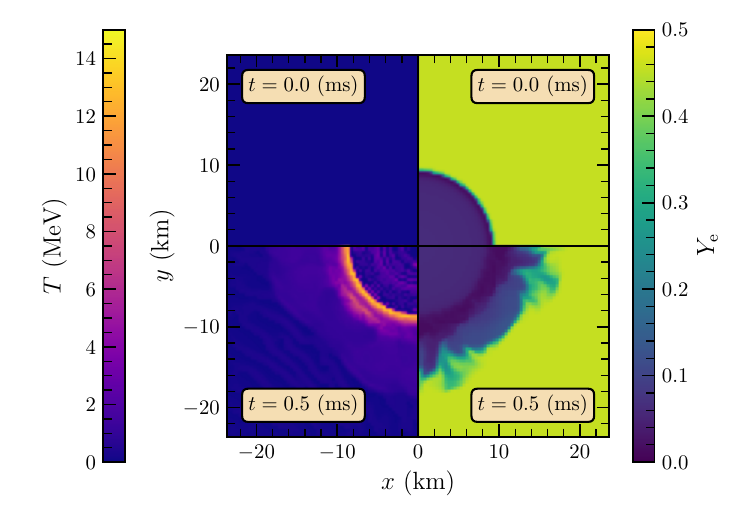}
    \caption{Temperature $T$ (left) and electron fraction $Y_\mathrm{e}$ 
             (right) of the initial configuration (top) and after 
             $0.5\ (\rm{ms})$ of evolution (bottom) of a $M = 1.4\ \Mo$ 
             TOV star using the SFHo \citep{Steiner:2012rk} EOS.}
 \label{fig:tabul_prim}
\end{figure}

In Figure~\ref{fig:tabul_prim} we show the temperature, $T$, and 
electron fraction, $Y_\mathrm{e}$, of the fluid at the initial timestep, 
and after $\sim 10$ light crossing times of the star 
($\sim 0.5\ \rm{ms}$). 
For the temperature we see the constant $T=0.1\ \rm{MeV}$ of the initial data at $t=0$,
however after the evolution we see some heating on the surface of the 
star, causing it to reach $T\approx 15\ \rm{MeV}$. This effect (seen in 
many other simulations, e.g.~\citep{Perego:2019adq, Endrizzi:2019trv, Prakash:2021wpz}) is driven by artificial shock 
heating as the sound speed falls off towards the edge of the star \citep{Gundlach:2010gy}. 
Additionally, we see a smaller increase in temperature in the core of 
the star, reaching around $T \approx 5\ \rm{MeV}$ in places. This effect 
comes from the poor condition number 
of the temperature with respect to 
the conserved energy $\tau$ \citep{Hammond:2021vtv}: small errors in $\tau$ due to, 
for example, interpolation of the spherically symmetric initial data 
onto the cartesian grid used in the evolution can be amplified by many 
orders of magnitude into errors in $T$ at high densities and low 
temperatures as the total energy of the fluid is very weakly dependent on $T$ 
under these conditions. For the electron fraction, we see at $t=0$
the expected equilibrium configuration of the electron 
fraction: a low $Y_\mathrm{e} \approx 0.05-0.1$ core surrounded by a 
higher $Y_\mathrm{e} \approx 0.45-0.5$ atmosphere. After the evolution we see 
that some of the low $Y_\mathrm{e}$ matter has begun to boil off the 
surface of the star, which, as mentioned in Section 
\ref{sec:ss:rns_grhd}, is a common artefact seen in simulations of this 
kind. The increase we see in Figure~\ref{fig:tabul_cons} is likely due 
to this effect.

 \subsection{Long-term Magnetic field dynamics in Static Neutron Star}\label{sec:ss:sns_grhmd}

The dynamics and re-configuration of an initially poloidal magnetic
field in a static NS remains an open problem in neutron star
physics. A poloidal magnetic field configuration is unstable on Alfven
timescales
\eg~ \citep{Tayler:1957a,Tayler:1973a,Wright:1973a,Markey:1973a,Markey:1974a,Flowers:1977a,Braithwaite:2005md}
and understanding magnetic field topology and energy 
re-distribution requires long-term numerical relativity evolutions,
\eg~\citep{Kiuchi:2008ss,Lasky:2011un,Ciolfi:2011xa,Pili:2017yxd,Sur:2021awe}. 

Here, we consider GRMHD evolutions of model A0 with a superposed
poloidal magnetic field given by the vector potential~\citep{Liu:2008xy} 
\begin{subequations}
\begin{align}
	&(A_x,A_y,A_z) = (-yA_\varphi, x A_\varphi, 0) \\
        &A_\varphi := A_b \max(p-0.04\, p_{\rm max},0) \label{eq:Apotential}
\end{align}
\end{subequations}
where $p_{\mathrm{max}}$ is the maximum value of the pressure within
the star. The parameter $A_b$ controls the magnitude of the magnetic
field, and is set to obtain a maximum value of $1.84\times10^{16}$~G
inside the star.
We use the WENOZ scheme and the same grid setup as for the model A0 in
Section~\ref{sec:ss:sns_grhd}. 
We observe similar features for the performance of the code in the case of GRMHD
as in the above GRHD simulations. We do
not repeat here the discussion and instead, discuss GRMHD effects
following \cite{Sur:2021awe}. 
In contrast to the simulations presented
in \cite{Sur:2021awe} which were performed in the Cowling approximation, we
now evolve the full dynamical spacetime. Due to the lack of implemented constraint 
preserving boundary conditions for the Z4c equations, we must adopt a different 
grid configuration to our previous simulations, with outer boundaries pushed 
further away. Consequently we use the same grid set up as in Section~\ref{sec:ss},
with resolutions $N=32,64,128$ closely matching the resolution over the NS itself
in simulations pS64, pS128, pS256 respectively in \cite{Sur:2021awe}.

 \begin{figure}[t]
   \centering 
     \includegraphics[width=0.49\textwidth]{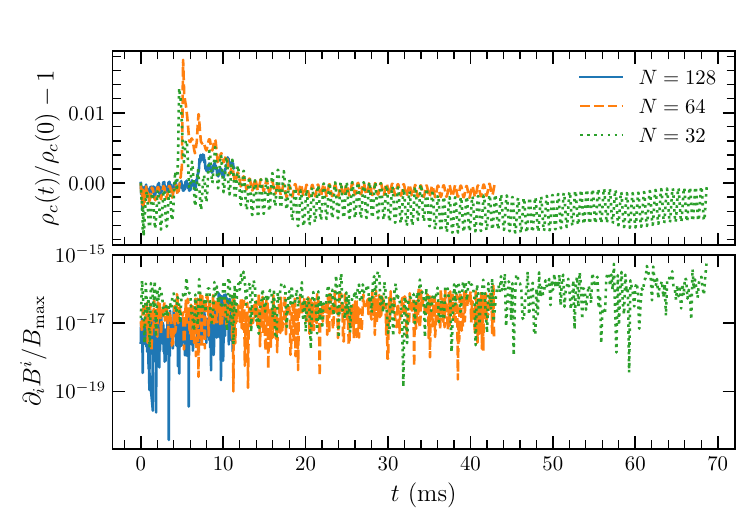} 
     \caption{Top: Evolution of central rest-mass density of the static
       star A0 model with superposed poloidal magnetic field with WENOZ reconstruction.
      Bottom: Integrated divergence of magnetic field (Eq.~\eqref{eq:gausslaw}) 
      normalised by maximum value of the magnetic field.
}
  \label{fig:A0_mhd_rhoc_divb}
 \end{figure}

In the upper panel of Figure~\ref{fig:A0_mhd_rhoc_divb} we see the oscillations of
the central density of the star. Excited by the superposed magnetic field these are larger 
in magnitude than those in Figure~\ref{fig:A0_rhoc}, with a peak in the oscillations 
corresponding to the saturation of the growth of $b^\phi$ as seen below.

The constrained transport algorithm is designed to preserve the
divergence-free condition on the magnetic field on the face
centred grid. As a first check, we show this is indeed
the case for our simulations. The lower panel of Figure~\ref{fig:A0_mhd_rhoc_divb} shows the 
integrated value of the divergence-free constraint over time at
different resolutions as normalised by the maximum value of $|B|$. 
The total relative violations oscillate at the level
of ${{\sim}10^{-16}}$ at the lowest resolution.

\begin{figure*}[t]
   \centering 
     \includegraphics[width=0.32\textwidth]{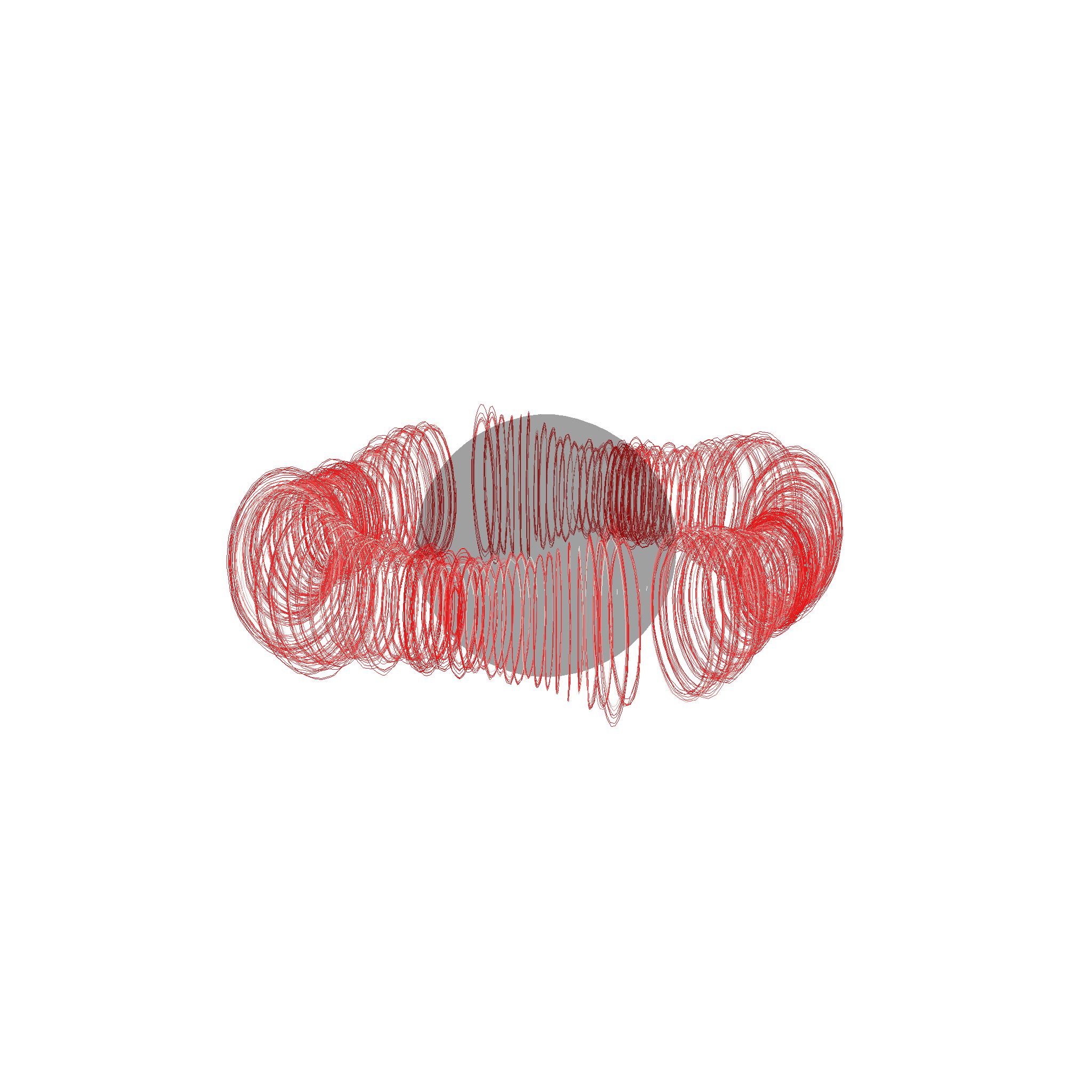}
     \includegraphics[width=0.32\textwidth]{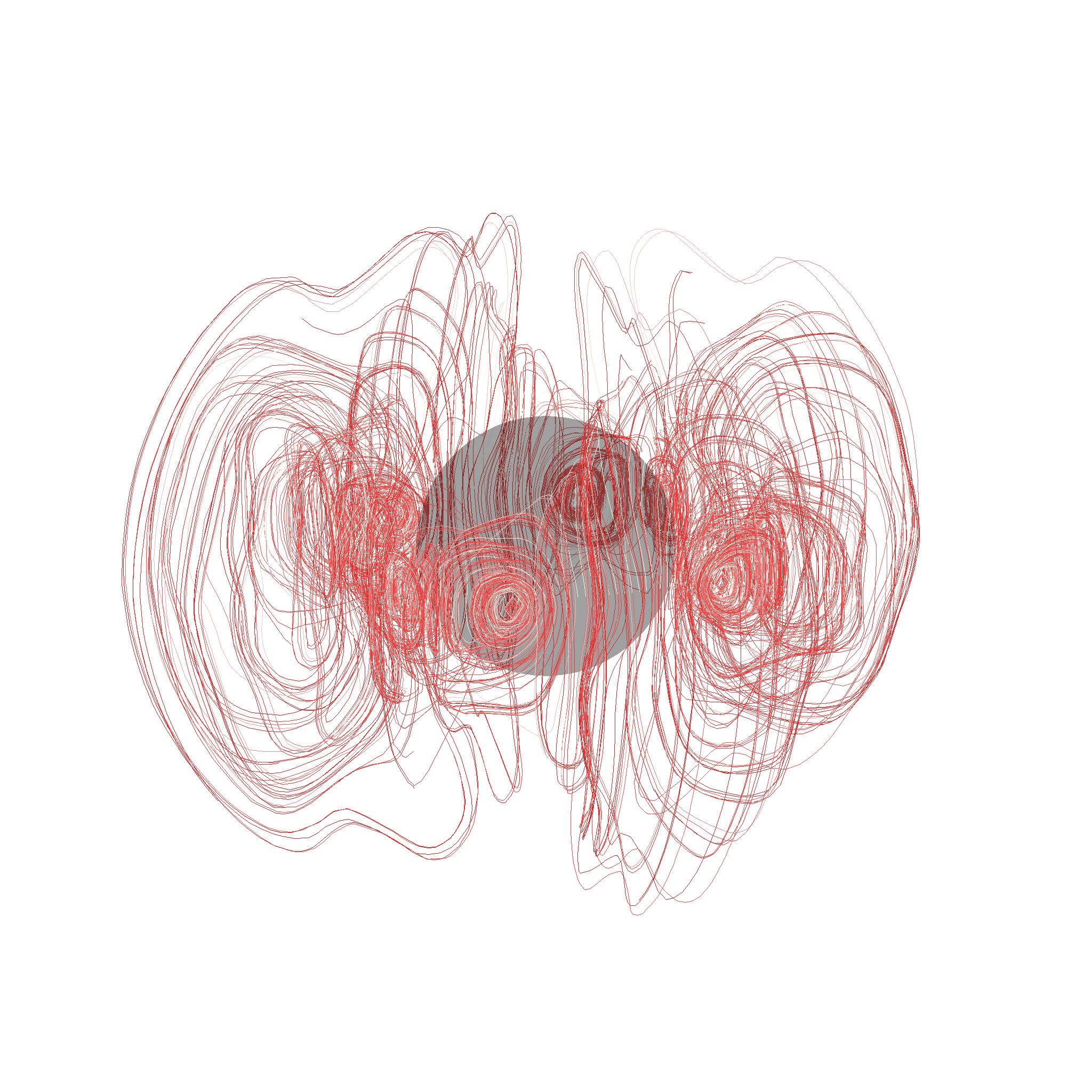}
     \includegraphics[width=0.32\textwidth]{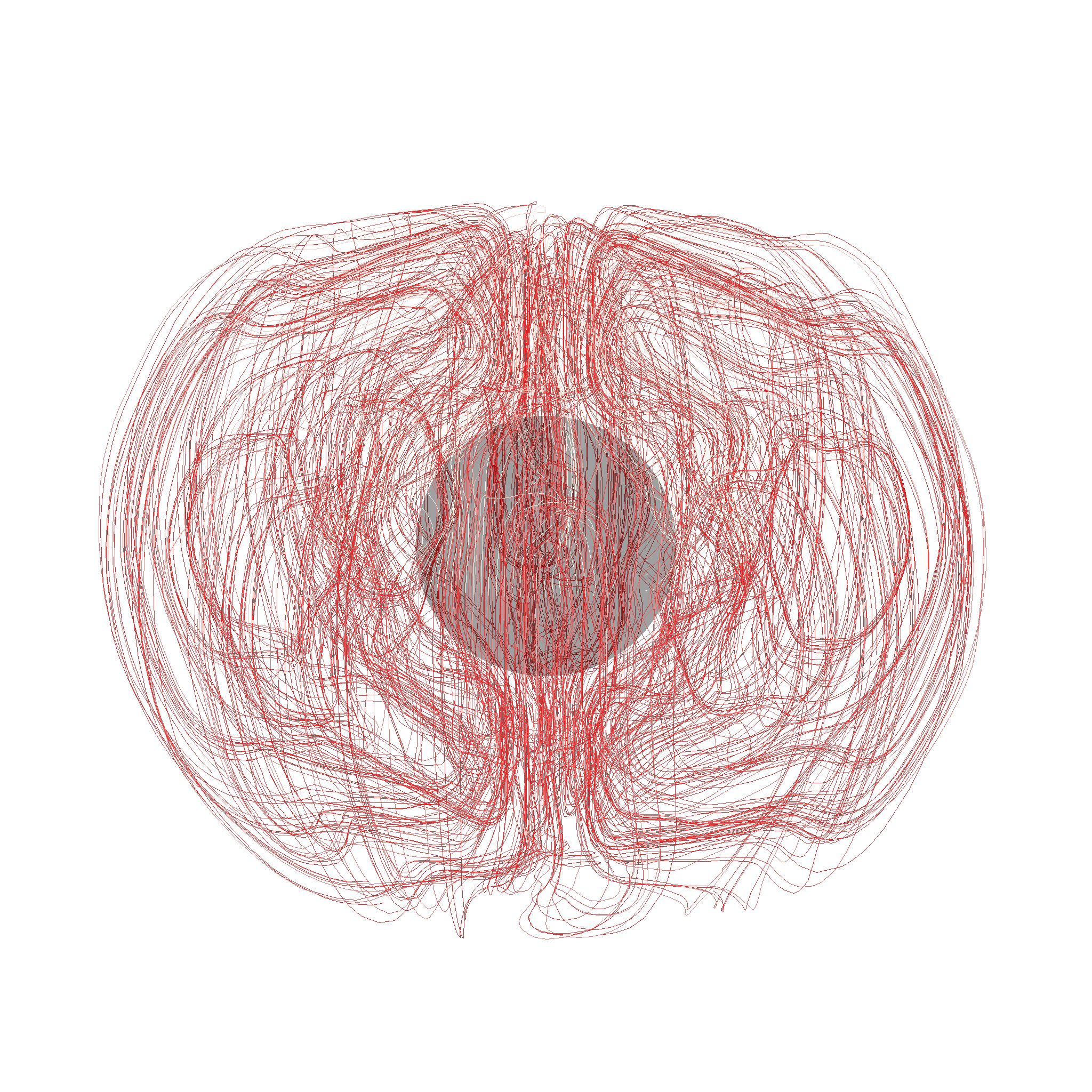}
     \caption{Evolution of magnetic streamlines seeded in equatorial plane at
       radius $5.91$ km. Gray isocontour shows surface of $\rho = 6.18\times 10^{14} \gccm$.
       Left: Varicose instability at $t=4.93$~ms.
       Middle: Kink instability at $t=12.8$~ms.
       Right: Late time non-linear field arrangement $t=44.3$~ms.}
  \label{fig:A0_mhd_streamline}
 \end{figure*}

We observe the onset of the well known ``varicose'' and ``kink'' instabilities
by visualising the evolution of magnetic field streamlines in the equatorial plane.
In Figure~\ref{fig:A0_mhd_streamline} we track the evolution of streamlines seeded on
a circle in the equatorial plane of radius $5.91$ km. After $4.93$ ms (left panel) we see the rotational
invariance of the magnetic streamlines is broken, with the onset of the varicose instability
altering the cross sectional area of these streamlines. As the evolution progresses further
we see some signs that the streamlines are further disrupted orthogonal to the equatorial 
plane in the onset of the kink instability, which we identify with the saturation of the growth in $b^\phi$ (see below),
 as can be seen by $12.8$ ms (middle panel). We finally show the late time appearance of the field after
its non-linear rearrangement, at $44.3$ ms (right panel). Here while we see an overall poloidal structure,
the field configuration has grown clear toroidal components.

\begin{figure}[t]
   \centering 
     \includegraphics[width=0.49\textwidth]{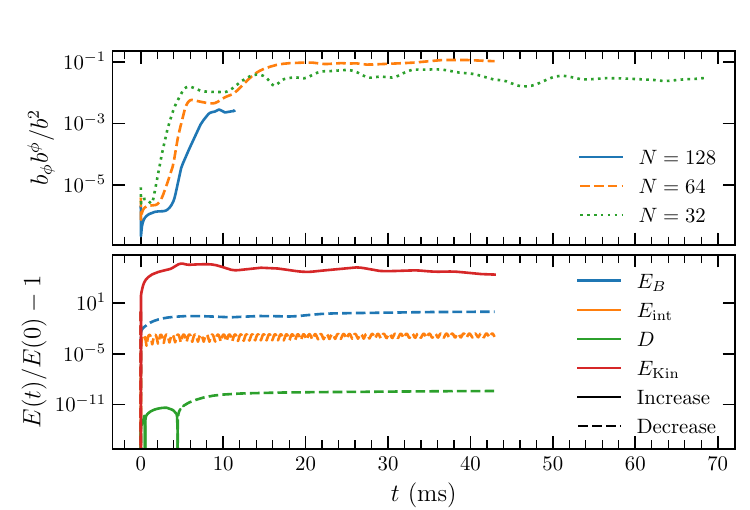} 
     \caption{Top: Growth of the toroidal component of magnetic field energy 
      as a fraction of total magnetic field energy. Bottom: relative change in 
      magnetic energy, internal energy, mass and kinetic energy for $N=64$. Solid lines denote 
      a positive change in energy, dashed lines denote a negative change.}
  \label{fig:A0_mhd_E}
 \end{figure}

The initially poloidal magnetic field is unstable, and so we observe the growth
of the toroidal component of the magnetic field, measured by $b^\phi b_\phi$. 
In the upper panel of Figure~\ref{fig:A0_mhd_E} we see that, after $\sim{}16$ ms, the toroidal
component grows to form $\sim{}10 \%$ of the total magnetic field energy, with 
the onset of the growth of this component delayed as a function of resolution.
This behaviour matches previously observed behaviour in the
Cowling approximation, demonstrated in Figure 3 of \cite{Sur:2021awe}.

In the lower panel of Figure~\ref{fig:A0_mhd_E} we track the relative changes 
in the energies arising from kinetic energy $E_{\mathrm{Kin}}$, magnetic energy $E_B$, 
internal energy $E_{\mathrm{int}}$ and the rest mass $D$. We see that as the magnetic field
instabilities drive the rearrangement of the field, the kinetic energy of the fluid correspondingly increases,
with local maxima in the kinetic energy corresponding to local maxima in the growing toroidal field, 
corresponding also to a loss in overall magnetic energy. We note also that the internal energy 
is well preserved in this simulation, in contrast to the enthalpy which was seen to grow in 
Figure 5 of \cite{Sur:2021awe}. We attribute this to an improved atmosphere treatment, 
and the increased grid size in comparison to the previous simulations.

\section{Gravitational collapse of Rotating Neutron Star}\label{sec:collapse}

In this section we benchmark \GRAthena{} against the gravitational
collapse of a rotating NS to a BH.  
Initial data are the unstable, uniformly rotating D4 equilibrium
configuration previously studied in several works,
\eg~\citep{Baiotti:2004wn,Reisswig:2012nc,Dietrich:2014wja}. The D4 model is spinning
close to the mass-shedding limit at a frequency of $1276$~Hz and
it has baryon mass and gravitational mass of, respectively,
$M_b=2.0443\Mo$ $M=1.8604\Mo$. 
The initial data are calculated using the \texttt{RNS} code and
imposing a central density of 
$\rho_c = 1.924388\times10^{15}\gccm$ 
and a polar-to-equatorial coordinate axis ratio of $r_p/r_e =
0.65$. Collapse is triggered by adding a $0.5\%$ perturbation to the
fluid pressure as in \cite{Dietrich:2014wja}.

We perform simulations at multiple grid resolutions for the PPM
reconstruction method. The grid is
composed of eight static refinement levels with $N=32,64,128$
and reaching a maximum resolution of $92.2,46.1,23.1$~m respectively.
The outer boundaries of the grid are the same as in Section~\ref{sec:ss:sns_grhd},
while the innermost 3 refinement levels are located within
the initial radius of the star.
In contrast to previous work we do not employ any grid symmetry.
Both reconstructions successfully handle BH 
formation and the
subsequent evolution of the latter without the use of excision.

 \begin{figure}[t]
   \centering 
     \includegraphics[width=0.49\textwidth]{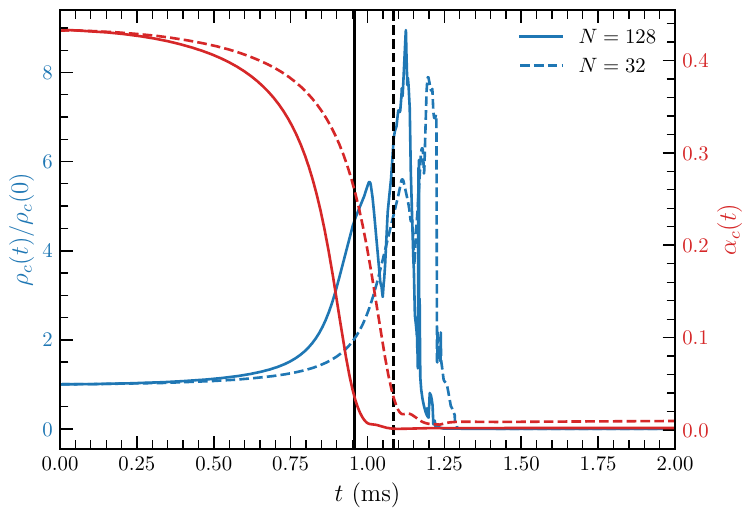}
     \caption{Evolution of central rest-mass density and lapse of
       collapsing D4 model. Data from PPM simulations are shown at
       different resolutions. Vertical black lines denote collapse
       time for each simulation. We note that the oscillatory behaviour occurs 
       within the horizon, after collapse.
     }
     \label{fig:D4_rhoc_alpc}
 \end{figure}

\begin{figure}[t]
   \centering 
   \includegraphics[width=0.49\textwidth]{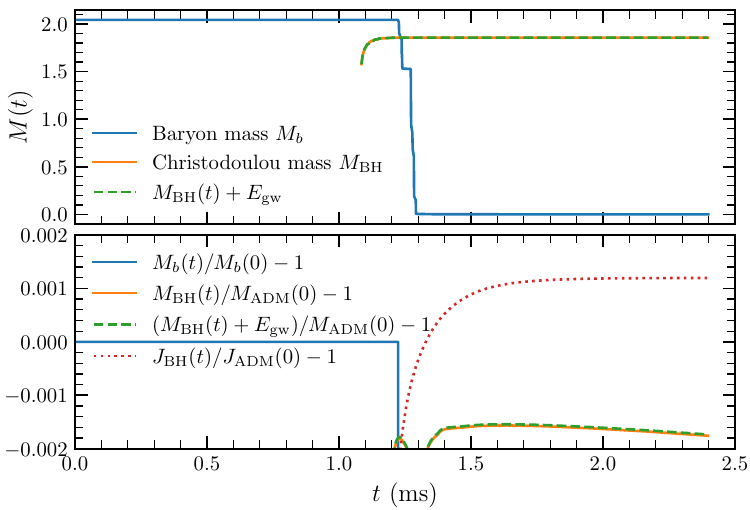}
   \caption{Top: Evolution of baryon mass and black hole (Christodoulou)
     mass in collapsing D4 model for $N=32$. The dashed line shows the
     sum of the black hole mass and the gravitational-wave energy in
     the $(2,0)$ mode extracted at the coordinate sphere.
     Bottom: Baryon mass and angular momentum conservation and
     relative differences of Christodoulou mass and with respect to
     the initial ADM mass.
   }
  \label{fig:D4_masses}
 \end{figure}

Figure~\ref{fig:D4_rhoc_alpc} shows the central density and lapse
during the collapse. As the density increases toward the central
region, the curvature increases and the lapse collapses as a result of
the gauge conditions. Black hole formation is handled by the moving
puncture gauge conditions adopted for the simulations
\citep{Thierfelder:2010dv}. 
For this simulation we employ no excision of the hydrodynamical variables
 within the horizon, in contrast to \eg~\citep{Bernuzzi:2020txg}, with mass 
loss from the grid occurring as the determinant of the spatial metric blows 
up at the origin.
An apparent horizon~\footnote{%
We implemented in \GRAthena{} an apparent horizon finder based on
(and following closely) the spectral fast-flow algorithm of
\cite{Gundlach:1997us,Alcubierre:1998rq}. Details will be given
elsewhere.}
(AH) is found at about one millisecond from the start
of the simulation. Until this point, the run shows excellent mass
conservation with a relative error of $10^{-13}$ up to collapse. We note that, since the initial
grid configuration contains refinement levels within the star, this provides 
a robust test of the mass conservation provided by the flux correction operators
described in \citep{Stone:2020}.
Afterwards, baryon mass is lost as shown in Figure~\ref{fig:D4_masses}.
The AH returns a Christodoulou mass \citep{Christodoulou:1970wf} of $M_{\rm BH}=1.857\Mo$ and an angular momentum $J_{\rm BH}=1.884\Mo^2$G/c. These values are
affected by relative uncertainties of order $10^{-3}$ for the lowest resolution 
simulation $N=32$, which improves as a function of resolution. These
results favourably compare to the results presented in
\citep{Reisswig:2012nc,Dietrich:2014wja}. 

\begin{figure}[t]
   \centering
   \includegraphics[width=0.49\textwidth]{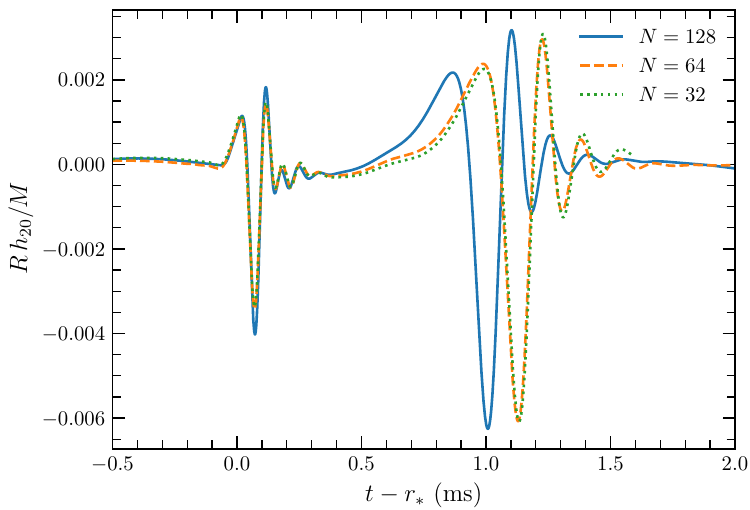}
   \caption{Gravitational waveforms of collapsing D4 model.
     Data from PPM simulations are shown at all resolutions for the
     dominant $(\ell,m)=(2,0)$ multipole of the strain. The waveform
     is shown as a function of the retarded time to the extraction
     sphere. At $N=32$ the resolution is too low to observe convergent behaviour, 
     though at higher resolutions we do see convergence in the waveforms at 
     approximate first order.
Note the y-axis is plotted in geometric units.
   }
   \label{fig:D4_waves}
 \end{figure}

Finally, we calculate the gravitational waveforms associated to the
collapse. Figure~\ref{fig:D4_waves} shows the $(\ell,m)=(2,0)$
multipole of the strain reconstructed from the Weyl scalar
\citep{Daszuta:2021ecf} using a time integration and corrected by a
polynomial drift \citep{Baiotti:2008nf}. Waveforms are extracted at
coordinate spheres centered at the origin of the Cartesian grid, constructed by sampling 
geodesic spheres, and of
radius $R=221$~km.
They are shown as a function of the retarded time to the extraction
spheres, given by $t-r_*$ with 
$r_* = r + 2 M \log(r/2M- 1)$, and $r(R)$ the areal radius of the
spheres of coordinate radius $R$ (the isotropic Schwarzschild radius).

After an initial burst of physically spurious so-called ``junk'' radiation,
the waveform has the
well-known ``precursor-burst-ringdown'' morphology expected for the
collapse. The amplitude's relative maximum at around $t-r_*\simeq 1$~ms corresponds
to AH formation, while the amplitude's absolute maximum is produced after BH
formation.
The black hole's quasi normal ringing follows the waveform
peak and is consistent with a perturbed Kerr metric \citep{Dietrich:2014wja}.
Overall, these tests indicate \GRAthena{} can handle gravitational
collapse and delivers reliable results at relatively low resolutions
and without grid symmetries.

\section{Binary Neutron Star spacetimes}\label{sec:bns}

This section presents tests that involve BNS spacetimes.
We discuss simulations of a quasi-circular, equal-mass merger with
and without magnetic fields. Irrotational constraint-satisfying
initial data are generated for a binary of baryon mass
$M_b=3.2500\Mo$ and gravitational mass $M=3.0297\Mo$ at an
initial separation of $45$~km. The Arnowitt-Deser-Misner (ADM) mass of
the binary is $M_{\rm ADM}=2.9984\Mo$, the angular momentum
$J_{\rm ADM}=8.83542\Mo^2$G/c, and the initial orbital frequency is
$f_0\simeq294$~Hz. 
These data are computed with the
\texttt{Lorene} library \citep{Gourgoulhon:2000nn} and are publicly
available~\footnote{%
Dataset \texttt{G2\_I14vs14\_D4R33\_45km} at
\url{https://lorene.obspm.fr/}.
}.
Here we use the ideal gas EOS with $K=124$.
For magnetised binaries, the magnetic field in each star is initialised as in the
single star case with the vector potential in Eq.~\eqref{eq:Apotential}, with $A_b$ 
chosen to give a maximum value of $1.77\times 10^{15} G$.
We first benchmark GRHD evolutions and
GWs against similar results obtained with the
\BAM
code~\citep{Bruegmann:1996kz,Brugmann:2008zz,Thierfelder:2011yi}.
We then focus on GRMHD evolutions and showcase the use of \GRAthena{}'s
AMR for investigating the Kelvin-Helmholtz instability developing at
the merger's collisional interface.

For all purely hydrodynamical runs described in this section, the grid outer 
boundaries are set at $[\pm 2268, \pm 2268, (0, 2268)]$~km, with
bitant symmetry enforced in the $z$ direction. For GRMHD simulations, 
no symmetry is enforced, and the outer boundaries are set at 
$[\pm 2268, \pm 2268, \pm2268]$~km. Initially a refined static 
grid is defined consisting of 7 levels of refinement, with the 
innermost grid covering both stars, extending from 
$[\pm 37, \pm 37, \pm 37]$~km, or $(0,37)$ km in the $z$ direction if
bitant symmetry is enforced across the $z=0$ plane. We perform runs at three base level resolutions,
$N=64,96,128$, corresponding to a grid spacing on the finest refinement
level of $(554,369,277)$~m respectively. 
Simulations are performed with the WENOZ reconstruction, and hydrodynamical variables
are excised within the apparent horizon for runs with GRMHD.

\subsection{Benchmark against \BAM} \label{sec:bns:vsBAM}

\begin{figure*}[t]
   \centering
    \includegraphics[width=0.32\textwidth]{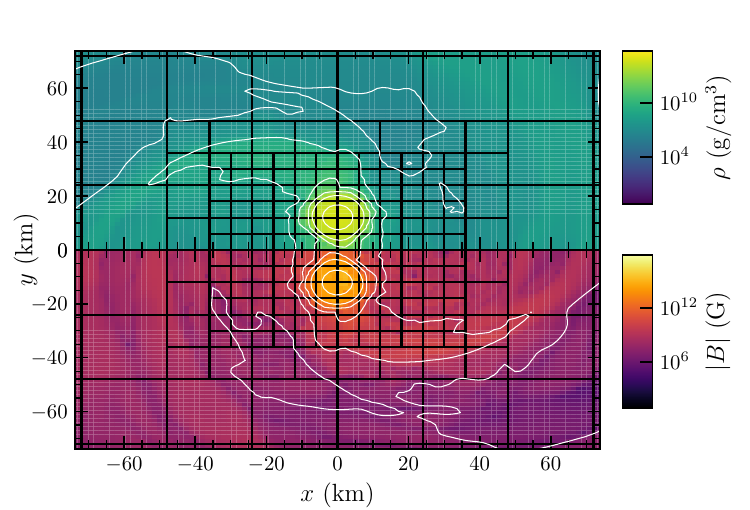}
    \includegraphics[width=0.32\textwidth]{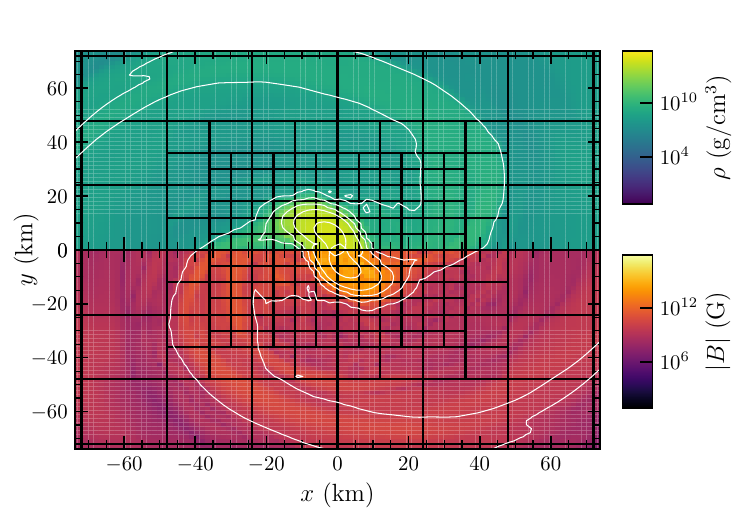}
    \includegraphics[width=0.32\textwidth]{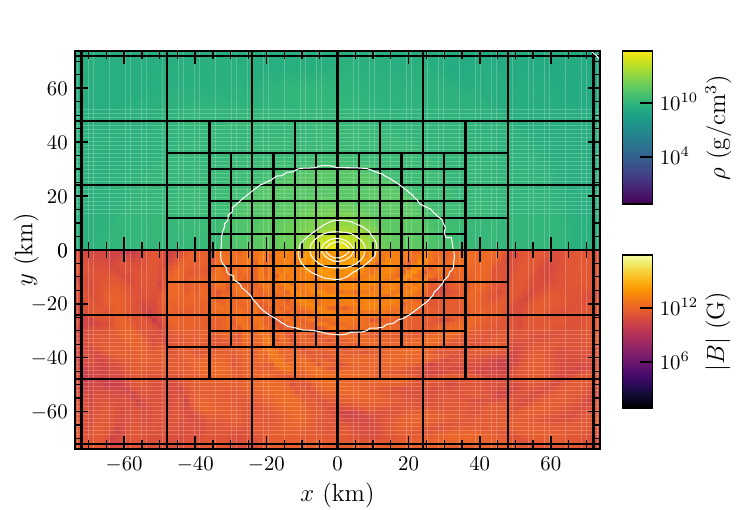}
     \caption{Snapshots of rest-mass density (upper half) and magnetic field strength
       (lower half)
       in the orbital plane. The snapshots correspond to the final orbit ($t=5.96$ms), the moment of
       merger ($t=7.83$ms), conventionally defined as the moment at which the
       $(\ell,m)=(2,2)$ mode of the gravitational wave has the
       amplitude peak, and a late stage shortly before gravitational collapse $t=16$ms. White contours are lines of constant density at values $\rho =(6.18 \times 10^{14}, 3.09\times 10^{14}, 6.18 \times 10^{13}, 6.18 \times 10^{12},6.18 \times 10^{10}, 6.18 \times 10^{8}, 6.18 \times 10^{6}, 6.18 \times 10^{4}) \gccm$  }
     \label{fig:bns_snap2d}
 \end{figure*}

\begin{figure}[t]
   \centering
     \includegraphics[width=0.49\textwidth]{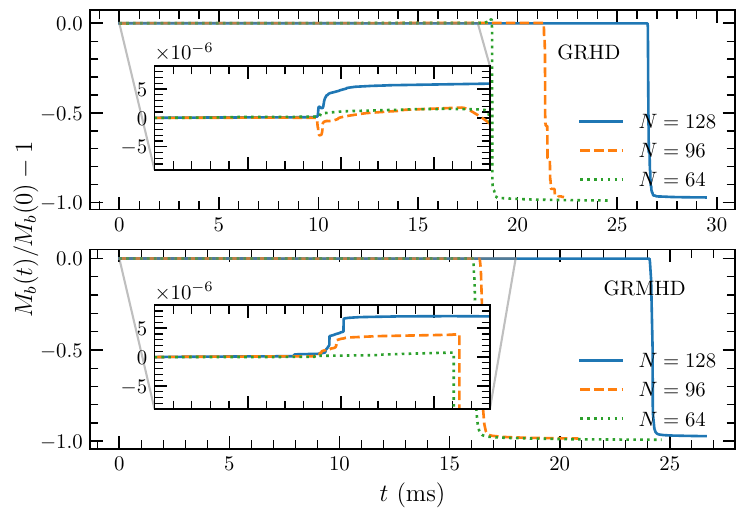}
     \caption{Baryon mass conservation along the GRHD and GRMHD
       evolutions of the binary neutron star model. To the moment
       of merger, $t\simeq7.5$~ms, the maximum relative violation is
       of the order $10^{-7}$. After merger, the violation increases to
       the $10^{-6}$  level. Note the mass typically increases
       rather than decreasing due to the accretion of
       the atmosphere. The sudden drop of the mass at late times is due to
       black hole formation.}
  \label{fig:BNS_masscons}
 \end{figure}

The binary revolves for about three orbits before forming a massive
remnant that undergoes gravitational collapse after about ${\sim}19$~ms
from the beginning of the simulation for the lowest resolution simulation with
GRHD.
We demonstrate 3 snapshots of the evolution of a magnetised
binary in Figure~\ref{fig:bns_snap2d}; at $t=5.96$~ms, shortly before merger; $t=7.83$~ms, 
approximately the moment of merger; and $t=16$~ms, shortly before gravitational
collapse to a BH, at which point an apparent horizon is found. We overlay the structure of the \Mesh{}, with each black 
square representing a single \MeshBlock{} of $16^3$  cells.
The baryon mass conservation is shown in Figure~\ref{fig:BNS_masscons} for
both the GRHD and GRMHD evolutions and for different resolutions. 
We find somewhat larger violations of mass conservation with respect to
the single star test, however, the maximum relative variation remains
at ${\sim}10^{-6}$ level at the lowest resolution. This value is
more than sufficient to robustly study mass ejecta and comparable to
other state-of-the-art Eulerian codes at the considered resolutions,
\eg~\citep{Radice:2018pdn}. Our experiments with the atmosphere
parameters  highlighted that simulations with the best mass conservation are obtained by
setups that lead to a mass increase (rather than decrease). 

The \GRAthena evolutions are in qualitative agreement with previous
\BAM simulations \citep{Thierfelder:2011yi}. For a quantitative
benchmark, we perform here two new \BAM simulations using six refinement
levels, two of which are moving following the stars. The finest
refinement level covers entirely the star at a resolution of
${\sim} 461$~m, that is comparable with \GRAthena $N=64$.
One \BAM simulation uses a finite volume method similar to \GRAthena
and based on the WENOZ reconstruction~\citep{Bernuzzi:2012ci}.
The other uses the entropy flux-limited (EFL) scheme of
\cite{Doulis:2022vkx}, which is a fifth-order accurate
finite-differencing scheme also making use of the WENOZ reconstruction. This run 
is conducted with the best hydrodynamics scheme available to date for
these binary simulations.

 \begin{figure}[t]
   \centering
   \includegraphics[width=0.49\textwidth]{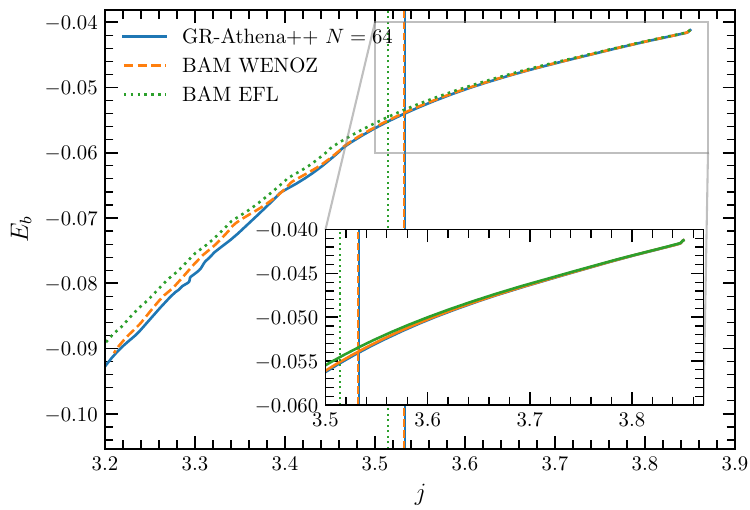}
     \caption{Binding energy \textit{vs} angular momentum curves
       $E_b(j)$ for \GRAthena and \BAM simulations.
       These quantities are computed following
       \cite{Damour:2011fu,Bernuzzi:2012ci}; they are rescaled by the mass,
       and shown in geometric units.       
       The moment of merger of each dataset is shown as vertical
       line. The inset zooms-in on the energetics during the orbital
       phase, prior to the moment of merger. Note the initial
       adjustment related to the use of conformally flat initial data
       and ``junk'' radiation.
     }     
  \label{fig:BNS_Ej}
 \end{figure}

A gauge-invariant description of the dynamics of binary spacetimes is
given by curves of binding energy and angular momentum
\citep{Damour:2011fu,Bernuzzi:2012ci}, which we use to compare evolutions.
In line with the above references we define the binding energy 
$E_b = (M_{\mathrm{ADM}} - E_{\mathrm{GW}})\nu/(M-1) $ and the angular 
momentum $j = (J_{\mathrm{ADM}} - J_{\mathrm{GW}})/(M^2\nu)$, 
with $\nu$ the symmetric mass ratio,
$\nu = q/(1+q)^2$, $q>1$ the mass ratio and quantities suffixed with 
$\mathrm{GW}$ the contributions radiated away in gravitational waves.

The energy curves for our
simulations are shown in Figure~\ref{fig:BNS_Ej}; 
note these quantities are 
commonly shown in geometric units. The evolution
proceeds from right to left, as the binary loses angular momentum and
becomes more bound. The agreement between \GRAthena and \BAM is
excellent, already at these low resolutions. Differences between the
simulations are more evident towards the moment of merger (vertical
lines) and afterwards. As is clear from the comparison of the two \BAM
simulations, these differences are mainly due to the hydrodynamics
scheme and are in particular affected by the choice of the
reconstruction scheme. We note that a key difference between
\BAM and \GRAthena 's implementation of the hydrodynamics schemes 
comes in the choice of the reconstructed primitive variables. \GRAthena 
reconstructs the fluid pressure $p$, while \BAM reconstructs the specific
internal energy $\epsilon$. Experimentation with this reconstruction choice
in \GRAthena leads to noticeable variations in the evolution as expected,
though we find more robust performance in \GRAthena when reconstructing 
the pressure. \GRAthena also requires the intergrid interpolations discussed
in Section~\ref{sec:intergrid} absent in \BAM, which uses a cell centred grid
for evolving the Einstein equations. Finally the grid structures between these
two codes are different, with \BAM utilising a nested box-in-box style 
refinement approach, as opposed to the block-based oct-tree structure of \GRAthena.
 
\begin{figure*}[t]
   \centering
   \includegraphics[width=0.99\textwidth]{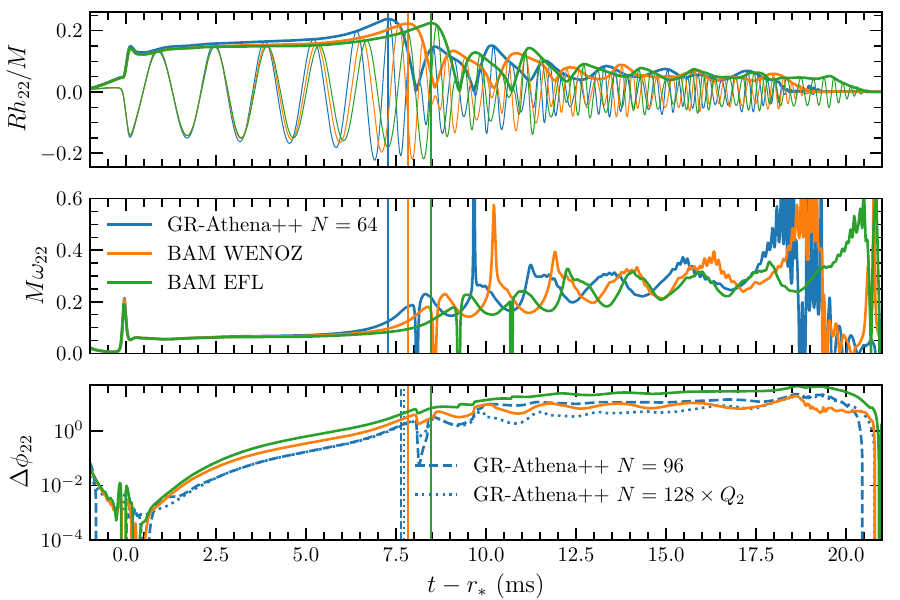}
   \caption{Waveform analysis and comparison with \BAM.
     Top: Real part and amplitude of the $(\ell,m)=(2,2)$ mode of the
     strain. We show \GRAthena data with WENOZ reconstruction and
     $N=64$ and \BAM data with finite volume scheme and WENOZ
     reconstruction and with the EFL scheme at a similar resolution.
     The y-axis is plotted in geometric units.
     Middle: Instantaneous frequency of the waves for the three datasets.
     The y-axis is plotted in geometric units.
     Bottom: phase differences between the \GRAthena data $N=64$ and
     \GRAthena data at higher resolution $N=96 $ (dashed) and $N=128$
     (dotted lines), and \BAM data. \GRAthena phase difference data at 
     $N=128$ is rescaled assuming second order convergence, matching with
     the $N=96$ data up to merger.
     The moment of merger of each dataset is shown as vertical line.
   }
  \label{fig:BNS_wvf}
 \end{figure*}

Finally, we compare the gravitational waveforms by discussing in
Figure~\ref{fig:BNS_wvf} the dominant $(\ell,m)=(2,2)$ mode of the
radiation. The different simulations mostly differ in the waveform
phase. At the lowest considered resolution the \GRAthena simulation
merges slightly earlier than the \BAM WENOZ and EFL.
This possibly indicates that the overall \GRAthena scheme and grid
choice leads to more numerical dissipation than the \BAM runs 
\citep{Bernuzzi:2012ci,Radice:2013hxh}, however, the time of collapse
of all three runs happens within a time interval of ${\sim}2.5$~ms,
indicating a substantial agreement of the computations.
The gravitational frequencies at the moment of merger (middle panel)
are very close to each other, $\omega_{22}\simeq 1340$~Hz for \GRAthena and \BAM WENOZ
and $\omega_{22}\simeq 1360$~Hz for \BAM EFL; as
are the remnants' frequency evolution. After collapse, the quasi
normal mode frequencies of the remnant black hole are poorly resolved
for all three runs at the considered resolution, but are compatible with 
the fundamental mode frequency $\nu_{\rm QNM}\sim6.5$~kHz
($M\omega_{22}\simeq0.57$ in geometric units).

The bottom panel of Figure~\ref{fig:BNS_wvf} quantifies the phase
differences of the \GRAthena $N=64$ run with respect to the \BAM runs
and the \GRAthena runs at higher resolutions. The difference between the $N=64$ and $N=128$
runs is rescaled according to Eq.~\eqref{eq:convfactor} to compare to the difference between the $N=64$ and $N=96$ run 
assuming second order convergence. These curves closely match up to merger suggesting 
second order convergence during the inspiral phase.
Further, the difference to the \BAM runs significantly reduces at higher
resolutions. 

\subsection{Kelvin-Helmholtz instability}\label{sec:bns:KH}

During the merger of a BNS the magnetic field in the initial NSs can be amplified
through a variety of instabilities triggered throughout the merger process. These
amplifications can lead to the large magnetic fields present in the remnant star
which may be required to launch relativistic jets that can be the source of the observed gamma 
ray bursts from the BNS merger associated to GW170817. One such instability is the Kelvin-Helmholtz instability (KHI), 
triggered when a shearing interface between the two stars is created, at the moment
of first contact between the stars. This creates small scale vortex like structures in the
fluid, associated to an amplification in the magnetic field \citep{Price:2006fi}.
Such amplifications have been extensively studied through numerical simulations
\citep{Kiuchi:2015sga,Kiuchi:2017zzg,Palenzuela:2021gdo,Aguilera-Miret:2023qih}.
Here we aim to use the flexible AMR of \GRAthena to efficiently resolve the 
KHI, with a targeted refinement criterion. Here we present the results of four simulations,
two with static grids set up as described in Section~\ref{sec:bns}, with resolution $N=64,128$, 
referred to here as SMR64 and SMR128, and two AMR runs initialised from the SMR64 run at $t=4.93$ms,  but allowed to refine 
by one further level, thus matching the resolution of the SMR128 run on the finest grid level. The
magnetic fields are initialised as in Section \ref{sec:bns}.
The first criterion for 
refinement is given by the maximum value of $|B|$ on the \MeshBlock{}, with a block 
refined if $|B_{\mathrm{max}}|$ exceeds $8.35 \times 10^{13}$G and derefined if lower than this value. 
We refer to this run as AMRB.
The second criterion is determined by the quantity
\begin{eqnarray}
\sigma = \sqrt{((\Delta_x \tilde u^y)^2 + (\Delta_y \tilde u^x)^2)},
\end{eqnarray}
with $\Delta_i$ the undivided difference operator in the $i$th coordinate direction, which monitors the shear of the fluid velocity.
\MeshBlock s are refined if $\sigma > 4.50 \times 10 ^9 cm/s$ anywhere within the block, and if the 
maximum density within the block exceeds a threshold value, in order to avoid overrefining
unphysical low-density regions, and they are derefined if $\sigma < 2.25 \times 10 ^9 cm/s$. We refer to this run as 
AMR$\sigma$. In both cases, 
derefinement is only permitted as long as the resolution does not drop below that of SMR64. 
Threshold values of $|B_{\mathrm{max}}|$ and $\sigma$ employed are determined by numerical experimentation.

\begin{figure*}[t]
   \centering
   \includegraphics[width=0.9\textwidth]{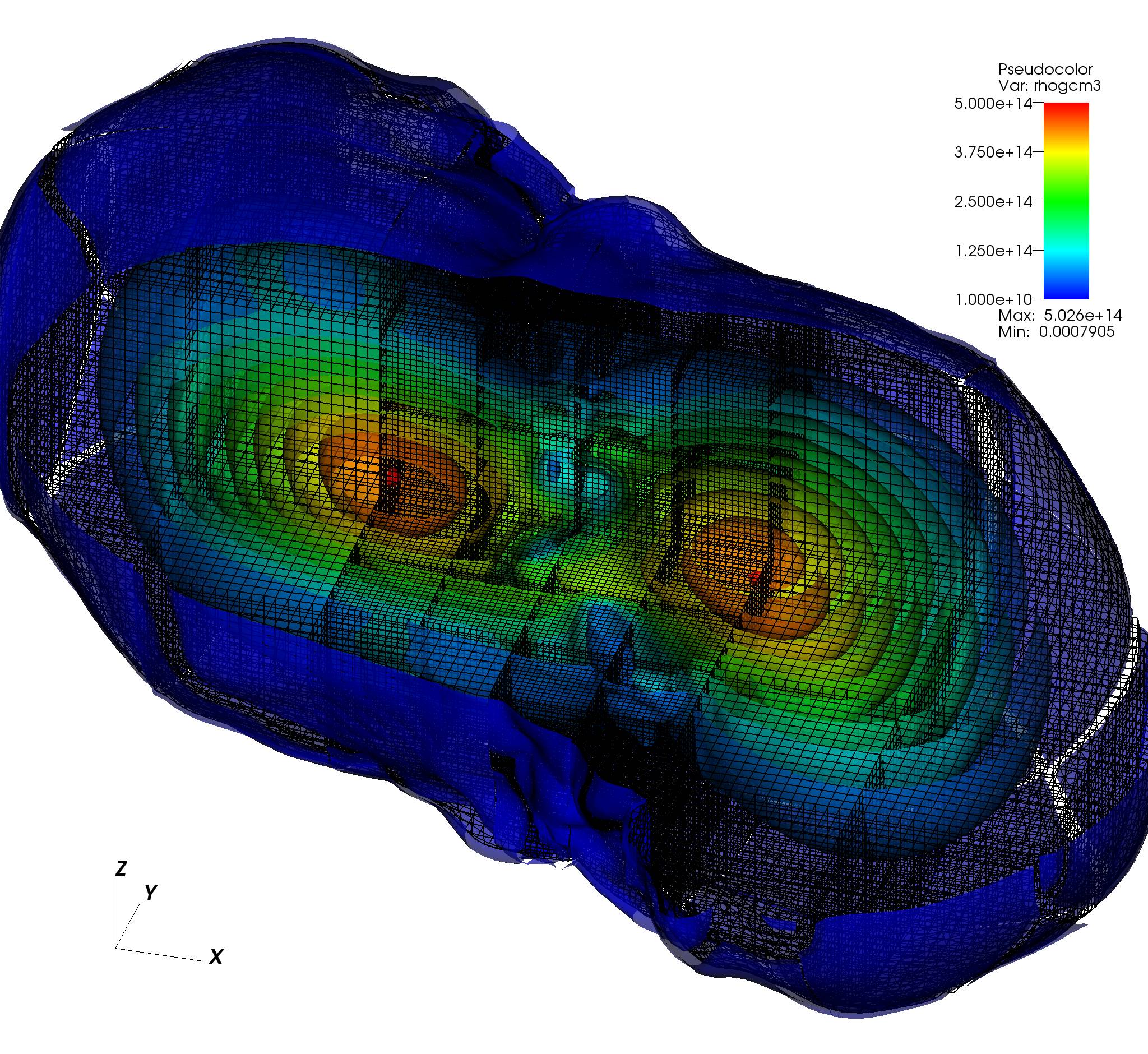}
   \caption{Snapshot of \Mesh{} structure at $t=7.88$ms shortly before merger in run AMR$\sigma$. Isocontours of density are also shown for $\rho > 10^{11} \gccm$. 
            For visualisation purposes the $z>0,~y<0$ quadrant has been cut out.
   }
  \label{fig:BNS_KHI_AMRgrid}
 \end{figure*}

We demonstrate the appearance of the grid evolution in Figure~\ref{fig:BNS_KHI_AMRgrid}, 
a snapshot at approximately the moment of merger at $t=7.88$ms of run AMR$\sigma$. 
Here the grid structure and density isocontours are shown for values of $\rho > 10^{11} \gccm$, with the $z>0,~y<0$ quadrant cut away for visualisation purposes.
We see that, at the interface between the stars, higher resolution \MeshBlock s have been generated in the region where the KHI driven 
amplification can be expected to occur.

To measure the effectiveness of the AMR criteria we demonstrate the amplification of the magnetic 
field $|B^2|$, as a measure of the magnetic energy, for the four runs described above, as well as the growth of the number of \MeshBlock s
in the AMR runs, which serves as a measure of computational cost. We note that, due to the global 
timestep of \Athena, as soon as the AMR runs create a block on the highest refinement level, the 
global timestep drops to match that of the run SMR128.

\begin{figure}[t]
   \centering
   \includegraphics[width=0.49\textwidth]{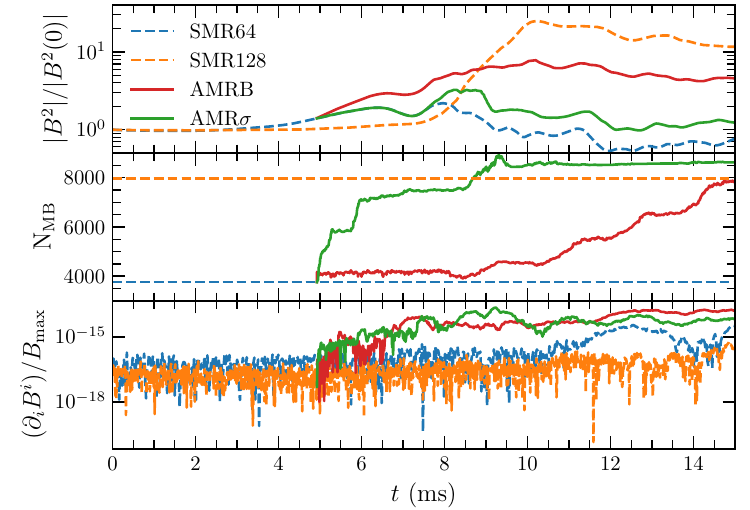}
   \caption{Top: Amplification of total magnetic field strength $|B^2|$ in BNS merger for SMR with 
   resolution $N=64,128$ and AMR criteria based on the maximum $B$ field (AMRB) and velocity shear 
   $\sigma$, (AMR$\sigma$). Middle: Total number of \MeshBlock s present in each run. 
   Bottom: $\partial_i B^i$ integrated over the computational domain, normalised by 
   the maximum value of the magnetic field.
   }
  \label{fig:BNS_KHI}
 \end{figure}

 In the upper panel of 
Figure~\ref{fig:BNS_KHI} we see the 
amplification of the magnetic field energy. For SMR128 we see that the 
energy is amplified by a factor of over 25 after the moment of merger, at $t\sim10.2~$ms,
through the KHI.
The AMR simulations manage to capture some of this amplification, with AMRB
 capturing an amplification of a factor of 7.85, while AMR$\sigma$ only captures a factor of 3.26. 
 We note that \cite{Kiuchi:2015sga}, at considerably higher resolutions of $17.5$m see a much larger 
amplification, of 6 orders of magnitude, from a much weaker initial magnetic field profile, of initial strength order $10^{13}$G. We expect that the addition of further levels of mesh refinement may be able 
to capture such larger amplifications, as well as the possibility of improved performance through
fine-tuning of the threshold values used in the AMR criteria. 

We also demonstrate the number of \MeshBlock s generated in each simulation, as a measure
for the computational cost, in the middle panel of Figure~\ref{fig:BNS_KHI}. 
We see that the run AMRB generates considerably fewer \MeshBlock s than 
the run SMR128 during the amplification of the field, with a factor of 1.79 fewer \MeshBlock s at the time of the peak in magnetic energy,
and a factor 1.91 fewer at the moment of merger itself. This suggests that targeted use of AMR 
in \GRAthena can allow us to resolve small features without drastically increasing the computational
cost, in contrast to a SMR grid configuration. In contrast we see that the criterion based 
on $\sigma$ captures less of the magnetic field amplification, and generates more \MeshBlock s than 
the SMR128 run. This criterion seems less suited to purely capturing the magnetic field amplification, but 
generates extra \MeshBlock s in lower density regions, at the star surface, and in the vicinity of
matter disrupted from the binary system. In this simulation however, we see superior conservation 
of mass, compared to both the SMR and other AMR runs, and note that such an AMR criterion may 
prove useful in tracking ejected matter for simulations with microphysical EOSs.

In the lower panel of Figure~\ref{fig:BNS_KHI}
we see the violation of the divergence-free condition on the magnetic field, Eq.~\eqref{eq:gausslaw}, integrated over the computational domain, as normalised by the maximum value of $|B|$, similar to 
the lower panel of Figure~\ref{fig:A0_mhd_rhoc_divb}.
Through the AMR runs we see at worst a total relative error of $10^{-14}$ in this constraint. This value is
larger by an order of magnitude than those for the SMR runs, but is maintained at a low level 
even though \MeshBlock s are 
continually created and destroyed as can be seen in the middle panel, verifying that
the divergence and curl preserving restriction and prolongation operators implemented in 
\citep{Stone:2020} of \citep{Toth:2002a}, are sufficient for the accurate performance of long-term 
BNS evolutions with magnetic fields and AMR.
 
\section{Scaling tests}\label{sec:scaling_tests}
In order to demonstrate the performance of \GRAthena{} on large problems
we perform scaling tests on several target machines, namely 
SuperMUC-NG (hereafter SuperMUC), HLRS-HAWK (hereafter HAWK) and Frontera,
running with both
OpenMP and MPI parallelizations. We consider SMR grid 
setups and different problems, namely a single TOV star and a
binary system consisting of two boosted TOV stars; both 
problems are run with and without magnetic fields.
When magnetic fields are included, the configuration matches 
that in Eq.~\eqref{eq:Apotential}.
Scaling performance is important for both problems;
for the binary problem we want to be able to efficiently run
at high resolution during the inspiral for the extraction of highly
accurate gravitational waveforms, while, at late times, high resolution
simulations of GRMHD processes in the post merger remnant require quality
scaling on a grid setup tuned to a single star.
In all the scaling tests presented $N_B = 16$, and we evolve 20
timesteps.

\subsection{Strong scaling tests}\label{ssec:strong_scaling}
The strong scaling tests are conducted as follows. We construct a
cubic SMR grid with the same physical extent as in Section~\ref{sec:bns}
with the same initial configuration of NSs.
We perform several strong 
scaling series by selecting different resolutions, 
allowing us to explore the scaling properties
at different numbers of cores.
Each series is then labelled by
$N$, which corresponds to a given resolution.
For each resolution the extent of the most refined region
is tuned in order to achieve similar \MeshBlock /core
ratios across all resolutions.
We present results obtained by utilising 
all the available cores in a node, namely using 48 cores 
(12 MPI tasks per node and 4 OpenMP threads per task)
for SuperMUC and 128 cores
(32 MPI tasks per node and 4 OpenMP threads per task) for HAWK.
We discuss the scaling properties, calculating the scaling efficiency as
\be\label{eq:efficiency}
\mathrm{~Efficiency~~} = 1 - \frac{t - t_{\mathrm{ideal}}}{t_{\mathrm{ideal}}},
\ee
where $t$ is the elapsed wallclock time and $t_\mathrm{ideal}$ is 
the time expected if the code had perfect strong scaling, halving exactly when resources 
were doubled.

 \begin{figure}[t]
   \centering
   \includegraphics[width=0.49\textwidth]{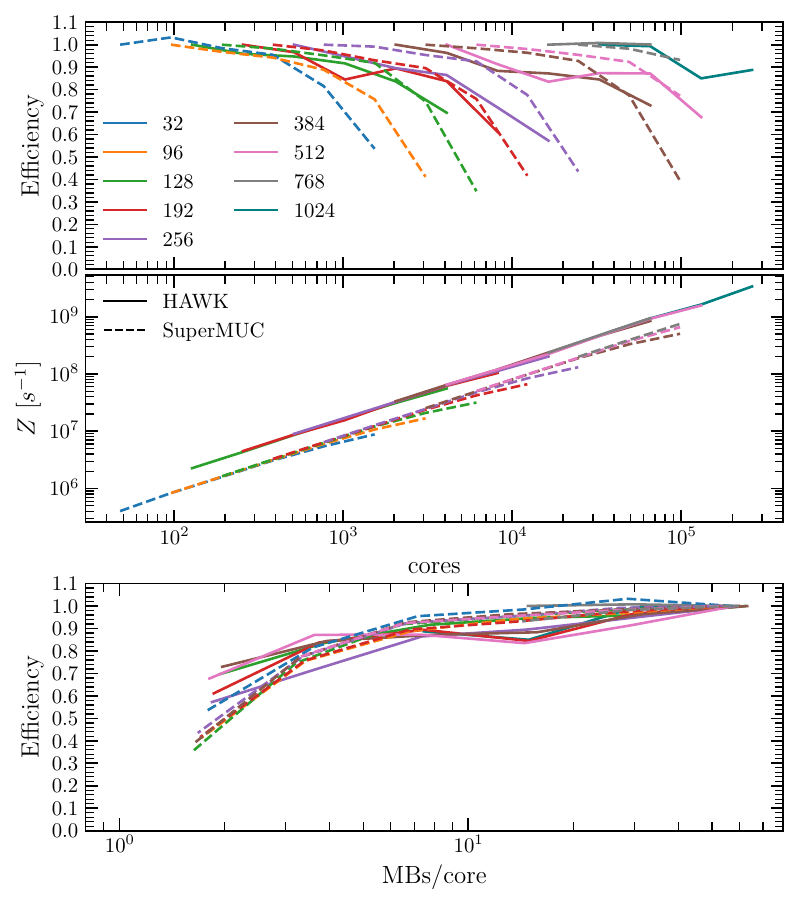}
   \caption{Scaling efficiency (see Eq. \eqref{eq:efficiency}) 
   for the binary neutron star configuration with MHD.
   Results for both HAWK and SuperMUC are included, up to
   $\mathcal{O}(10^5)$ cores. Top: efficiency versus number
   of cores. Middle: zone cycles per second ($Z$) versus number
   of cores.
   Bottom: efficiency in terms of the \MeshBlock s/core ratio.
   }
  \label{fig:binary_scaling}
 \end{figure}
In Figure \ref{fig:binary_scaling} we show the scaling 
efficiency calculated 
from our strong scaling tests 
for the magnetised binary, performed on
both SuperMUC and HAWK clusters. 
In the top panel we demonstrate very good strong scaling efficiencies 
for all resolutions, \ie in all regimes of number of cores. In particular,
we find efficiencies in excess of $80\%$ up to $\sim 10^5$ cores (gray and teal lines) 
for both SuperMUC (dashed lines) and HAWK (solid lines).
In the scaling tests on HAWK, for some resolutions (\eg pink and brown lines), 
we observe a faster initial drop in efficiency compared to the tests on SuperMUC.
After this drop, however, the efficiency settles at a constant value $\sim 85\%$ and decreases again
to $\sim 70 \%$ only when the load goes below 4 \MeshBlock s/core (see third panel). 
The bottom panel of Figure \ref{fig:binary_scaling} shows that the efficiency
on SuperMUC is $\gtrsim 90\%$ as long as $\mathrm{\MeshBlock s/core} \gtrsim 6$.
We attribute such different behaviours to the different architectures
of the two computer clusters. This causes a difference in the raw performance in terms of 
zone cycles per second ($Z$) (middle panel), where we note that runs on HAWK are a factor 
$\gtrsim 2$ faster than the runs on SuperMUC.

We note that we find comparable results also for the binary problem
without magnetic field evolution, and the single star evolution with and without magnetic fields.

\subsection{Weak scaling tests}\label{ssec:weak_scaling}

We measure the performance of weak scaling by measuring the zone cycles per second $Z$ performed
by the code, with the expectation that for perfect scaling, this should double every time the 
computational load and resources are concurrently doubled. We measure the efficiency of this process
by 

\be\label{eq:weakefficiency}
\mathrm{~Weak~ Efficiency~~} = 1 - \frac{Z - Z_{\mathrm{ideal}}}{Z_{\mathrm{ideal}}}.
\ee

\begin{figure}[t]
   \centering
     \includegraphics[width=0.49\textwidth]{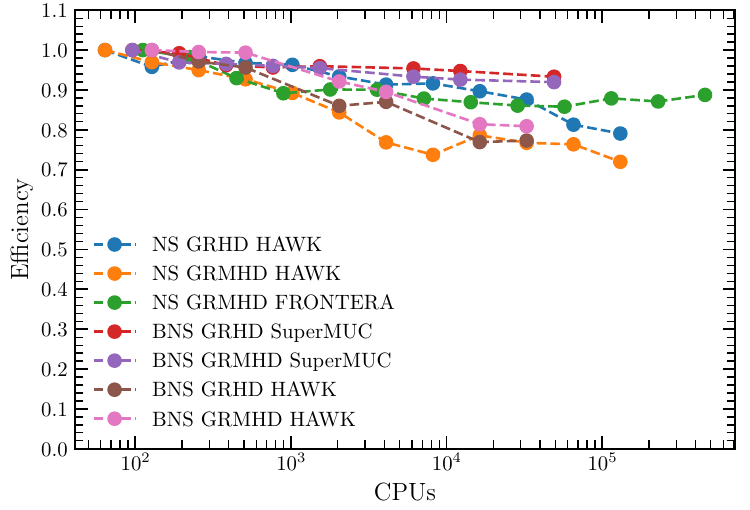}
     \caption{Weak scaling (Eq.~\eqref{eq:weakefficiency}) for a single star and a binary neutron star configuration (NS and BNS, respectively) with and without magnetic fields (GRHD and GRMHD, respectively)
      for grids with SMR, comparing results on different target machines.}
  \label{fig:scaling_weak}
 \end{figure}

In Figure~\ref{fig:scaling_weak} we show the weak scaling performance of \GRAthena for a variety of 
problems, on various machines. We test a single star both with and without magnetic fields on 
HAWK and Frontera, and perform binary tests on HAWK and SuperMUC. We see weak 
scaling maintained up to $\sim{}5\times 10^5$ CPU cores for the single star test at 89\% efficiency on 
Frontera, with efficiencies dropping to, at worst 72\% on HAWK. For binary tests we see an efficiency
$\gtrsim 90\%$ up to $\sim{}5\times 10^4$ cores on SuperMUC-NG, with slightly lower efficiencies 
seen on HAWK for the same problem. These results are consistent with our previous results
from the vacuum case of \cite{Daszuta:2021ecf} and those for stationary spacetimes in \cite{Stone:2020}.

\section{Conclusion}\label{sec:final_word}

This paper presents an extension of \GRAthena{} that allows us to
perform GRMHD simulations of astrophysical flows on dynamical
spacetimes. The code implements 3+1 Eulerian GRMHD in conservative
form and couples it to the Z4c metric solver presented in
\cite{Daszuta:2021ecf}. The MHD solver is a direct extension of \Athena{}'s
constrained-transport scheme 
\citep{Stone:2020}, implemented on an oct-tree mesh with block-based AMR. 
Other novel technical aspects introduced with respect to the original codebase 
are grid-to-grid operators between cell, vertex and face centered grid
variables, WENO reconstruction schemes, a generic EOS interface and a
new conservative-to-primitive solver.

The code is mainly targeted to astrophysical applications involving
neutron star spacetimes. We have demonstrated it can successfully pass a
standard set of challenging benchmarks, including
the long-term stable evolution of isolated equilibrium neutron
star configurations, 
 the gravitational collapse of an unstable rotating
neutron star to a black hole (and the subsequent black hole evolution),
and binary neutron star mergers from inspiral to merger
remnants. For the latter tests, we have demonstrated gravitational waveforms that
are convergent and consistent with the \BAM code.
Notably, most of the performed tests involve no grid symmetries
and rather low resolutions, while still providing quantitatively
correct results.

Anticipating full-scale applications, we have discussed novel simulations
of magnetic field instabilities in both isolated and merging neutron
stars. We have performed long-term evolutions of isolated magnetised 
neutron stars with initially poloidal fields up to 68.7 ms 
extending our previous simulations
in \cite{Sur:2021awe} to include full dynamical spacetime evolution. Here we find
similar results for the growth of a toroidal field component, saturating at approximately 
$10\%$ of the total magnetic field energy, while finding a superior conservation
of the internal energy of the neutron star, and total relative violations of the divergence-free
condition on the magnetic field of order $10^{-16}$.

We have also performed mergers of magnetised binary neutron stars utilising 
the full AMR capabilities of \Athena{} to efficiently resolve the Kelvin-Helmholtz 
instability, investigating the efficiency and performance of two refinement strategies
based on dynamically evolving field values. Even at initially low resolutions we find
an amplification of a factor $\sim{}8$ when refining by 1 extra level based on the 
magnitude of the magnetic field, while utilising a factor $\sim{}2$ fewer \MeshBlock s
at the moment of merger, compared to an SMR resolution of the same resolution at the 
NSs.

The scaling properties of \GRAthena{} have been tested on a variety of machines
and for differing problems, with strong scaling efficiencies in excess of $80\%$ shown
up to $10^5$ CPU cores, and weak scaling efficiency in excess of $90\%$ as far as our tests have extended, up to
 $\sim{}5\times 10^4$ CPU cores for a magnetised binary neutron star evolution. 
For a single star we also tested the scaling behaviour, with weak scaling holding 
at efficiencies of up to $89\%$ on $\sim{}5\times 10^5$ CPU cores. This scaling 
efficiency will allow us to tackle large GRMHD problems without symmetries. 

Work is ongoing on further improving some algorithmic aspects and 
physics modules of \GRAthena.
A code version employing cell-centered metric fields and high-order
restrict-prolong operators is also under testing. A future paper will
report on a detailed comparison of the performances of vertex-centered
and cell-centered metric representation for vacuum and neutron star
spacetimes.  
For the computation of accurate
gravitational waveforms we are implementing high-order schemes which
have proven to be essential for waveform convergence and 
overall quality \citep{Radice:2013hxh,Bernuzzi:2016pie,Doulis:2022vkx}.
Further, the accuracy of computations of
merger remnants and other strong-gravity phenomena 
depends crucially on the detailed simulation of microphysics.
To this aim, we are porting into \GRAthena{} the neutrino transport
scheme developed by \cite{Radice:2021jtw} together with a physically improved
implementation of weak reactions and reaction rates.
We also plan to couple to \GRAthena{} the sophisticated radiation solvers
recently developed in the \Athena{} framework by
\cite{Bhattacharyya:2022bzf,White:2023wxh}.

\acknowledgments

BD, SB acknowledge funding from 
the EU H2020 under ERC Starting Grant, no.~BinGraSp-714626, and 
from the EU Horizon under ERC Consolidator Grant,
no. InspiReM-101043372.
SB acknowledges funding from the Deutsche Forschungsgemeinschaft, DFG,
project MEMI number BE 6301/2-1.
PH acknowledges funding from the National Science Foundation under Grant No.~PHY-2116686.
DR acknowledges funding from the U.S. Department of Energy, Office of
Science, Division of Nuclear Physics under Award Number(s) DE-SC0021177,
from the National Science Foundation under Grants No. PHY-2011725,
PHY-2020275, PHY-2116686, and AST-2108467, and NASA under award
No.~80NSSC21K1720.
Simulations were performed on the ARA cluster at Friedrich Schiller
University Jena, on the supercomputer SuperMUC-NG at the Leibniz-
Rechenzentrum (LRZ, \url{www.lrz.de}) Munich, and on the national HPE
Apollo Hawk at the High Performance Computing Center Stuttgart (HLRS).
The ARA cluster is funded in part by DFG grants INST 275/334-1
FUGG and INST 275/363-1 FUGG, and ERC Starting
Grant, grant agreement no. BinGraSp-714626. 
The authors acknowledge the Gauss Centre for Supercomputing
e.V. (\url{www.gauss-centre.eu}) for funding this project by providing
computing time on the GCS Supercomputer SuperMUC-NG at LRZ
(allocations {\tt pn36ge} and {\tt pn36jo}).
The authors acknowledge HLRS for funding this project by pro-
viding access to the supercomputer HPE Apollo Hawk
under the grant numbers INTRHYGUE/44215 and GRA-BNS-Scaling/44244.
Simulations were also performed on Frontera at the Texas Advanced
Computing Center (TACC) and enabled by National Science Foundation LRAC
allocation PHY23001
This research used resources of the National Energy Research Scientific
Computing Center, a DOE Office of Science User Facility supported by the
Office of Science of the U.S.~Department of Energy under Contract
No.~DE-AC02-05CH11231.


\appendix

\section{\PrimitiveSolver}
\label{app:PrimitiveSolver}

\PrimitiveSolver is a new conserved-to-primitive library based on the algorithm of 
\citet{Kastaun:2020uxr} (hereafter KKC). However, it implements a small number of changes.

First, temperature is treated as a fundamental thermodynamic variable over the specific
internal energy, $\epsilon$. While the distinction is trivial for an ideal gas, this
convention coincides better with the nuclear EOS tables commonly employed in core-collapse
supernova and BNS simulations. In principle this also means that common thermodynamic
expressions which require expensive root solves on tables can now be written as lookup
operations, though in practice this is not always the case.

Secondly, internal \PrimitiveSolver calculations use the total energy density 
$e = \rho(1 + \epsilon)$ in place of $\epsilon$. This offers a small computational 
advantage, as the estimate of $e$ from the root solver, $\hat{e}$, requires fewer 
floating-point operations than the calculation of $\hat{\epsilon}$ as described by KKC,
without any significant round-off error for velocities near zero. Certain intermediate
calculations can also be carried through, resulting in approximately five fewer 
floating-point operations for a single root iteration. 

\PrimitiveSolver passes an extensive set of unit tests to demonstrate self-consistency
between the primitive-to-conserved and conserved-to-primitive procedures, typically
maintaining a relative error of $10^{-10}$ or less for all but the most extreme situations
($W\sim10^3$ for an ideal gas, $W\sim10^2$ for a tabulated EOS). We have additionally
validated our implementation against the reference implementation of KKC in \reprimand in
systematic benchmarks, and we find that our algorithmic changes do not result in any
noticeable difference in accuracy. We also find that the root solver in \PrimitiveSolver
is more robust and often converges faster, particularly for large $W$, than \reprimand.

The most important difference between \PrimitiveSolver and \reprimand, however, is
architectural: the latter utilizes a more traditional polymorphic design, with each EOS
inheriting from an abstract interface class and implementing specific EOS behavior through
virtual functions. \PrimitiveSolver instead uses a templated policy-based design; the base
EOS interface accepts a specific equation of state as a template parameter and inherits
directly from that class. While adding an element of complexity to the interface class,
this change eliminates the need for virtual functions, which has benefits for vectorizing
the code and porting it to other architectures, such as GPUs.

In addition to the EOS, \PrimitiveSolver also templates over the error policy, which 
includes both the atmosphere treatment and responses to problems occurring during the
primitive solve. The standard policy used in \GRAthena defines an atmosphere with 
$\rho = \rho_\textrm{atm}$, $T = T_\textrm{atm}$, and $v^i = 0$. If either $\rho$ or $D$
falls below $\rho_\textrm{atm}f_\textrm{thr}$, where $f_\textrm{thr}$ is a thresholding
coefficient, all hydrodynamical variables are reset to atmosphere. If $T$ falls below 
$T_\textrm{atm}$, $T$ is reset, and the pressure is recalculated. During the 
conserved-to-primitive inversion, a density and composition-dependent floor assuming 
$T=0$ is calculated for $\tau$. If $\tau$ falls below this value, it is reset, but all
other variables are left untouched.

Because the GRMHD equations become ill-conditioned when $B^2/D \gg 1$ or $W \gg 1$, limits
are imposed on these variables during the conserved-to-primitive inversion. If 
$B^2/D > M_\textrm{max}$, $D$ is rescaled to $B^2/M_\textrm{max}$. If
$W > W_\textrm{max}$, $W$ is reset to $W_\textrm{max}$ without modifying the conserved
variables, effectively increasing $\rho$.

Lastly, during the root solve step for the inversion procedure, it can happen that the
calculated values of $\rho$, $e$, or $T$ fall outside the bounds permitted by the EOS, 
particularly when using a tabulated EOS. In these circumstances, the quantity is simply
reset to the nearest permissible value.


\begin{thebibliography}{}
\expandafter\ifx\csname natexlab\endcsname\relax\def\natexlab#1{#1}\fi
\providecommand{\url}[1]{\href{#1}{#1}}
\providecommand{\dodoi}[1]{doi:~\href{http://doi.org/#1}{\nolinkurl{#1}}}
\providecommand{\doeprint}[1]{\href{http://ascl.net/#1}{\nolinkurl{http://ascl.net/#1}}}
\providecommand{\doarXiv}[1]{\href{https://arxiv.org/abs/#1}{\nolinkurl{https://arxiv.org/abs/#1}}}

\bibitem[{Abbott {et~al.}(2017{\natexlab{a}})}]{TheLIGOScientific:2017qsa}
Abbott, B.~P., {et~al.} 2017{\natexlab{a}}, Phys. Rev. Lett., 119, 161101,
  \dodoi{10.1103/PhysRevLett.119.161101}

\bibitem[{Abbott {et~al.}(2017{\natexlab{b}})}]{Monitor:2017mdv}
---. 2017{\natexlab{b}}, Astrophys. J., 848, L13,
  \dodoi{10.3847/2041-8213/aa920c}

\bibitem[{Abbott {et~al.}(2017{\natexlab{c}})}]{GBM:2017lvd}
---. 2017{\natexlab{c}}, Astrophys. J., 848, L12,
  \dodoi{10.3847/2041-8213/aa91c9}

\bibitem[{Aguilera-Miret {et~al.}(2023)Aguilera-Miret, Palenzuela, Carrasco, \&
  Vigan\`o}]{Aguilera-Miret:2023qih}
Aguilera-Miret, R., Palenzuela, C., Carrasco, F., \& Vigan\`o, D. 2023.
\newblock \doarXiv{2307.04837}

\bibitem[{Alcubierre {et~al.}(2000)Alcubierre, Brandt, Bruegmann, Gundlach,
  Masso, Seidel, \& Walker}]{Alcubierre:1998rq}
Alcubierre, M., Brandt, S., Bruegmann, B., {et~al.} 2000, Class. Quant. Grav.,
  17, 2159, \dodoi{10.1088/0264-9381/17/11/301}

\bibitem[{Alic {et~al.}(2012)Alic, Bona-Casas, Bona, Rezzolla, \&
  Palenzuela}]{Alic:2011gg}
Alic, D., Bona-Casas, C., Bona, C., Rezzolla, L., \& Palenzuela, C. 2012,
  Phys.Rev., D85, 064040, \dodoi{10.1103/PhysRevD.85.064040}

\bibitem[{Alic {et~al.}(2013)Alic, Kastaun, \& Rezzolla}]{Alic:2013xsa}
Alic, D., Kastaun, W., \& Rezzolla, L. 2013, Phys. Rev., D88, 064049,
  \dodoi{10.1103/PhysRevD.88.064049}

\bibitem[{Anderson {et~al.}(2006)Anderson, Hirschmann, Liebling, \&
  Neilsen}]{Anderson:2006ay}
Anderson, M., Hirschmann, E., Liebling, S.~L., \& Neilsen, D. 2006,
  Class.Quant.Grav., 23, 6503, \dodoi{10.1088/0264-9381/23/22/025}

\bibitem[{Anderson {et~al.}(2008)Anderson, Hirschmann, Lehner, Liebling, Motl,
  {et~al.}}]{Anderson:2007kz}
Anderson, M., Hirschmann, E.~W., Lehner, L., {et~al.} 2008, Phys.Rev., D77,
  024006, \dodoi{10.1103/PhysRevD.77.024006}

\bibitem[{{Anile}(1990)}]{Anile:1990a}
{Anile}, A.~M. 1990, {Relativistic Fluids and Magneto-fluids}

\bibitem[{Anninos {et~al.}(2005)Anninos, Fragile, \&
  Salmonson}]{Anninos:2005kc}
Anninos, P., Fragile, P.~C., \& Salmonson, J.~D. 2005, Astrophys. J., 635, 723,
  \dodoi{10.1086/497294}

\bibitem[{Anton {et~al.}(2006)Anton, Zanotti, Miralles, Marti, Ibanez,
  {et~al.}}]{Anton:2005gi}
Anton, L., Zanotti, O., Miralles, J.~A., {et~al.} 2006, Astrophys.J., 637, 296,
  \dodoi{10.1086/498238}

\bibitem[{Arcavi {et~al.}(2017)}]{Arcavi:2017xiz}
Arcavi, I., {et~al.} 2017, Nature, 551, 64, \dodoi{10.1038/nature24291}

\bibitem[{Baiotti {et~al.}(2009)Baiotti, Bernuzzi, Corvino, De~Pietri, \&
  Nagar}]{Baiotti:2008nf}
Baiotti, L., Bernuzzi, S., Corvino, G., De~Pietri, R., \& Nagar, A. 2009, Phys.
  Rev., D79, 024002, \dodoi{10.1103/PhysRevD.79.024002}

\bibitem[{Baiotti {et~al.}(2005)Baiotti, Hawke, Montero, Loffler, Rezzolla,
  {et~al.}}]{Baiotti:2004wn}
Baiotti, L., Hawke, I., Montero, P.~J., {et~al.} 2005, Phys.Rev., D71, 024035,
  \dodoi{10.1103/PhysRevD.71.024035}

\bibitem[{Baker {et~al.}(2006)Baker, Centrella, Choi, Koppitz, \& van
  Meter}]{Baker:2005vv}
Baker, J.~G., Centrella, J., Choi, D.-I., Koppitz, M., \& van Meter, J. 2006,
  Phys. Rev. Lett., 96, 111102, \dodoi{10.1103/PhysRevLett.96.111102}

\bibitem[{{Balbus} \& {Hawley}(1991)}]{Balbus:1991a}
{Balbus}, S.~A., \& {Hawley}, J.~F. 1991, \apj, 376, 214,
  \dodoi{10.1086/170270}

\bibitem[{{Balsara}(1998)}]{Balsara:1998b}
{Balsara}, D.~S. 1998, \apjs, 116, 133, \dodoi{10.1086/313093}

\bibitem[{Banyuls {et~al.}(1997)Banyuls, Font, Ibanez, Marti, \&
  Miralles}]{Banyuls:1997zz}
Banyuls, F., Font, J.~A., Ibanez, J. M.~A., Marti, J. M.~A., \& Miralles, J.~A.
  1997, Astrophys. J., 476, 221

\bibitem[{Baumgarte \& Shapiro(1999)}]{Baumgarte:1998te}
Baumgarte, T.~W., \& Shapiro, S.~L. 1999, Phys. Rev., D59, 024007,
  \dodoi{10.1103/PhysRevD.59.024007}

\bibitem[{Beckwith \& Stone(2011)}]{Beckwith:2011iy}
Beckwith, K., \& Stone, J.~M. 2011, Astrophys. J. Suppl., 193, 6,
  \dodoi{10.1088/0067-0049/193/1/6}

\bibitem[{Bernuzzi \& Dietrich(2016)}]{Bernuzzi:2016pie}
Bernuzzi, S., \& Dietrich, T. 2016, Phys. Rev., D94, 064062,
  \dodoi{10.1103/PhysRevD.94.064062}

\bibitem[{Bernuzzi \& Hilditch(2010)}]{Bernuzzi:2009ex}
Bernuzzi, S., \& Hilditch, D. 2010, Phys. Rev., D81, 084003,
  \dodoi{10.1103/PhysRevD.81.084003}

\bibitem[{Bernuzzi {et~al.}(2012)Bernuzzi, Nagar, Thierfelder, \&
  Br{\"u}gmann}]{Bernuzzi:2012ci}
Bernuzzi, S., Nagar, A., Thierfelder, M., \& Br{\"u}gmann, B. 2012, Phys.Rev.,
  D86, 044030, \dodoi{10.1103/PhysRevD.86.044030}

\bibitem[{Bernuzzi {et~al.}(2020)}]{Bernuzzi:2020txg}
Bernuzzi, S., {et~al.} 2020, Mon. Not. Roy. Astron. Soc.,
  \dodoi{10.1093/mnras/staa1860}

\bibitem[{Bhattacharyya \& Radice(2022)}]{Bhattacharyya:2022bzf}
Bhattacharyya, M.~K., \& Radice, D. 2022, \dodoi{10.1016/j.jcp.2023.112365}

\bibitem[{{Blinnikov} {et~al.}(1984){Blinnikov}, {Novikov}, {Perevodchikova},
  \& {Polnarev}}]{Blinnikov:1984a}
{Blinnikov}, S.~I., {Novikov}, I.~D., {Perevodchikova}, T.~V., \& {Polnarev},
  A.~G. 1984, Soviet Astronomy Letters, 10, 177,
  \dodoi{10.48550/arXiv.1808.05287}

\bibitem[{Borges {et~al.}(2008)Borges, Carmona, Costa, \& Don}]{Borges:2008a}
Borges, R., Carmona, M., Costa, B., \& Don, W.~S. 2008, Journal of
  Computational Physics, 227, 3191, \dodoi{10.1016/j.jcp.2007.11.038}

\bibitem[{Braithwaite \& Spruit(2006)}]{Braithwaite:2005md}
Braithwaite, J., \& Spruit, H.~C. 2006, Astron. Astrophys., 450, 1097,
  \dodoi{10.1051/0004-6361:20041981}

\bibitem[{Brandt \& Br{\"u}gmann(1997)}]{Brandt:1997tf}
Brandt, S., \& Br{\"u}gmann, B. 1997, Phys. Rev. Lett., 78, 3606,
  \dodoi{10.1103/PhysRevLett.78.3606}

\bibitem[{Br{\"u}gmann(1996)}]{Bruegmann:1996kz}
Br{\"u}gmann, B. 1996, Phys. Rev., D54, 7361, \dodoi{10.1103/PhysRevD.54.7361}

\bibitem[{Br{\"u}gmann {et~al.}(2008)Br{\"u}gmann, Gonzalez, Hannam, Husa,
  Sperhake, {et~al.}}]{Brugmann:2008zz}
Br{\"u}gmann, B., Gonzalez, J.~A., Hannam, M., {et~al.} 2008, Phys.Rev., D77,
  024027, \dodoi{10.1103/PhysRevD.77.024027}

\bibitem[{Campanelli {et~al.}(2006)Campanelli, Lousto, Marronetti, \&
  Zlochower}]{Campanelli:2005dd}
Campanelli, M., Lousto, C.~O., Marronetti, P., \& Zlochower, Y. 2006, Phys.
  Rev. Lett., 96, 111101, \dodoi{10.1103/PhysRevLett.96.111101}

\bibitem[{Cheong {et~al.}(2021)Cheong, Lam, Ng, \& Li}]{Cheong:2020kpv}
Cheong, P. C.-K., Lam, A. T.-L., Ng, H. H.-Y., \& Li, T. G.~F. 2021, Mon. Not.
  Roy. Astron. Soc., 508, 2279, \dodoi{10.1093/mnras/stab2606}

\bibitem[{Christodoulou(1970)}]{Christodoulou:1970wf}
Christodoulou, D. 1970, Phys. Rev. Lett., 25, 1596,
  \dodoi{10.1103/PhysRevLett.25.1596}

\bibitem[{Ciolfi(2018)}]{Ciolfi:2018tal}
Ciolfi, R. 2018, Int. J. Mod. Phys. D, 27, 1842004,
  \dodoi{10.1142/S021827181842004X}

\bibitem[{Ciolfi {et~al.}(2011)Ciolfi, Lander, Manca, \&
  Rezzolla}]{Ciolfi:2011xa}
Ciolfi, R., Lander, S.~K., Manca, G.~M., \& Rezzolla, L. 2011, Astrophys.J.,
  736, L6, \dodoi{10.1088/2041-8205/736/1/L6}

\bibitem[{Ciolfi \& Rezzolla(2013)}]{Ciolfi:2013dta}
Ciolfi, R., \& Rezzolla, L. 2013, Mon. Not. Roy. Astron. Soc., 435, L43,
  \dodoi{10.1093/mnrasl/slt092}

\bibitem[{Cipolletta {et~al.}(2020)Cipolletta, Kalinani, Giacomazzo, \&
  Ciolfi}]{Cipolletta:2019geh}
Cipolletta, F., Kalinani, J.~V., Giacomazzo, B., \& Ciolfi, R. 2020, Class.
  Quant. Grav., 37, 135010, \dodoi{10.1088/1361-6382/ab8be8}

\bibitem[{{Colella} {et~al.}(2011){Colella}, {Dorr}, {Hittinger}, \&
  {Martin}}]{Colella:2011a}
{Colella}, P., {Dorr}, M.~R., {Hittinger}, J.~A.~F., \& {Martin}, D.~F. 2011,
  Journal of Computational Physics, 230, 2952,
  \dodoi{10.1016/j.jcp.2010.12.044}

\bibitem[{Combi \& Siegel(2023{\natexlab{a}})}]{Combi:2022nhg}
Combi, L., \& Siegel, D.~M. 2023{\natexlab{a}}, Astrophys. J., 944, 28,
  \dodoi{10.3847/1538-4357/acac29}

\bibitem[{Combi \& Siegel(2023{\natexlab{b}})}]{Combi:2023yav}
---. 2023{\natexlab{b}}.
\newblock \doarXiv{2303.12284}

\bibitem[{Coulter {et~al.}(2017)}]{Coulter:2017wya}
Coulter, D.~A., {et~al.} 2017, Science, \dodoi{10.1126/science.aap9811}

\bibitem[{Curtis {et~al.}(2022)Curtis, M\"osta, Wu, Radice, Roberts,
  Ricigliano, \& Perego}]{Curtis:2021guz}
Curtis, S., M\"osta, P., Wu, Z., {et~al.} 2022, Mon. Not. Roy. Astron. Soc.,
  518, 5313, \dodoi{10.1093/mnras/stac3128}

\bibitem[{Damour {et~al.}(2012)Damour, Nagar, Pollney, \&
  Reisswig}]{Damour:2011fu}
Damour, T., Nagar, A., Pollney, D., \& Reisswig, C. 2012, Phys.Rev.Lett., 108,
  131101, \dodoi{10.1103/PhysRevLett.108.131101}

\bibitem[{Daszuta {et~al.}(2021)Daszuta, Zappa, Cook, Radice, Bernuzzi, \&
  Morozova}]{Daszuta:2021ecf}
Daszuta, B., Zappa, F., Cook, W., {et~al.} 2021, Astrophys. J. Supp., 257, 25,
  \dodoi{10.3847/1538-4365/ac157b}

\bibitem[{de~Haas {et~al.}(2022)de~Haas, Bosch, M\"osta, Curtis, \&
  Schut}]{deHaas:2022ytm}
de~Haas, S., Bosch, P., M\"osta, P., Curtis, S., \& Schut, N. 2022.
\newblock \doarXiv{2208.05330}

\bibitem[{{Dedner} {et~al.}(2002){Dedner}, {Kemm}, {Kr{\"o}ner}, {Munz},
  {Schnitzer}, \& {Wesenberg}}]{Dedner:2002a}
{Dedner}, A., {Kemm}, F., {Kr{\"o}ner}, D., {et~al.} 2002, Journal of
  Computational Physics, 175, 645, \dodoi{10.1006/jcph.2001.6961}

\bibitem[{Del~Zanna {et~al.}(2007)Del~Zanna, Zanotti, Bucciantini, \&
  Londrillo}]{DelZanna:2007pk}
Del~Zanna, L., Zanotti, O., Bucciantini, N., \& Londrillo, P. 2007, Astron.
  Astrophys., 473, 11, \dodoi{10.1051/0004-6361:20077093}

\bibitem[{Dietrich \& Bernuzzi(2015)}]{Dietrich:2014wja}
Dietrich, T., \& Bernuzzi, S. 2015, Phys.Rev., D91, 044039,
  \dodoi{10.1103/PhysRevD.91.044039}

\bibitem[{Dimmelmeier {et~al.}(2006)Dimmelmeier, Stergioulas, \&
  Font}]{Dimmelmeier:2005zk}
Dimmelmeier, H., Stergioulas, N., \& Font, J.~A. 2006, Mon. Not. Roy. Astron.
  Soc., 368, 1609, \dodoi{10.1111/j.1365-2966.2006.10274.x}

\bibitem[{Doulis {et~al.}(2022)Doulis, Atteneder, Bernuzzi, \&
  Br\"ugmann}]{Doulis:2022vkx}
Doulis, G., Atteneder, F., Bernuzzi, S., \& Br\"ugmann, B. 2022, Phys. Rev. D,
  106, 024001, \dodoi{10.1103/PhysRevD.106.024001}

\bibitem[{Duez {et~al.}(2008)Duez, Foucart, Kidder, Pfeiffer, Scheel, \&
  Teukolsky}]{Duez:2008rb}
Duez, M.~D., Foucart, F., Kidder, L.~E., {et~al.} 2008, Phys. Rev., D78,
  104015, \dodoi{10.1103/PhysRevD.78.104015}

\bibitem[{Duez {et~al.}(2005)Duez, Liu, Shapiro, \& Stephens}]{Duez:2005sf}
Duez, M.~D., Liu, Y.~T., Shapiro, S.~L., \& Stephens, B.~C. 2005, Phys. Rev.,
  D72, 024028, \dodoi{10.1103/PhysRevD.72.024028}

\bibitem[{Duez {et~al.}(2003)Duez, Marronetti, Shapiro, \&
  Baumgarte}]{Duez:2002bn}
Duez, M.~D., Marronetti, P., Shapiro, S.~L., \& Baumgarte, T.~W. 2003, Phys.
  Rev., D67, 024004, \dodoi{10.1103/PhysRevD.67.024004}

\bibitem[{East {et~al.}(2012)East, Pretorius, \& Stephens}]{East:2011aa}
East, W.~E., Pretorius, F., \& Stephens, B.~C. 2012, Phys.Rev., D85, 124010,
  \dodoi{10.1103/PhysRevD.85.124010}

\bibitem[{Eichler {et~al.}(1989)Eichler, Livio, Piran, \&
  Schramm}]{Eichler:1989ve}
Eichler, D., Livio, M., Piran, T., \& Schramm, D.~N. 1989, Nature, 340, 126,
  \dodoi{10.1038/340126a0}

\bibitem[{Endrizzi {et~al.}(2020)Endrizzi, Perego, Fabbri, Branca, Radice,
  Bernuzzi, Giacomazzo, Pederiva, \& Lovato}]{Endrizzi:2019trv}
Endrizzi, A., Perego, A., Fabbri, F.~M., {et~al.} 2020, Eur. Phys. J. A, 56,
  15, \dodoi{10.1140/epja/s10050-019-00018-6}

\bibitem[{Etienne {et~al.}(2010)Etienne, Liu, \& Shapiro}]{Etienne:2010ui}
Etienne, Z.~B., Liu, Y.~T., \& Shapiro, S.~L. 2010, Phys. Rev., D82, 084031,
  \dodoi{10.1103/PhysRevD.82.084031}

\bibitem[{Etienne {et~al.}(2015)Etienne, Paschalidis, Haas, M\"osta, \&
  Shapiro}]{Etienne:2015cea}
Etienne, Z.~B., Paschalidis, V., Haas, R., M\"osta, P., \& Shapiro, S.~L. 2015,
  Class. Quant. Grav., 32, 175009, \dodoi{10.1088/0264-9381/32/17/175009}

\bibitem[{Etienne {et~al.}(2012)Etienne, Paschalidis, Liu, \&
  Shapiro}]{Etienne:2011re}
Etienne, Z.~B., Paschalidis, V., Liu, Y.~T., \& Shapiro, S.~L. 2012, Phys.Rev.,
  D85, 024013, \dodoi{10.1103/PhysRevD.85.024013}

\bibitem[{{Evans} \& {Hawley}(1988)}]{Evans:1988a}
{Evans}, C.~R., \& {Hawley}, J.~F. 1988, Astrophys. J., 332, 659,
  \dodoi{10.1086/166684}

\bibitem[{{Flowers} \& {Ruderman}(1977)}]{Flowers:1977a}
{Flowers}, E., \& {Ruderman}, M.~A. 1977, \apj, 215, 302,
  \dodoi{10.1086/155359}

\bibitem[{Font(2007)}]{Font:2007zz}
Font, J.~A. 2007, Living Rev. Rel., 11, 7

\bibitem[{Font {et~al.}(2000)Font, Miller, Suen, \& Tobias}]{Font:1998hf}
Font, J.~A., Miller, M.~A., Suen, W.-M., \& Tobias, M. 2000, Phys. Rev., D61,
  044011

\bibitem[{Font {et~al.}(2002)}]{Font:2001ew}
Font, J.~A., {et~al.} 2002, Phys. Rev., D65, 084024,
  \dodoi{10.1103/PhysRevD.65.084024}

\bibitem[{Foucart {et~al.}(2011)Foucart, Duez, Kidder, \&
  Teukolsky}]{Foucart:2010eq}
Foucart, F., Duez, M.~D., Kidder, L.~E., \& Teukolsky, S.~A. 2011, Phys. Rev.,
  D83, 024005, \dodoi{10.1103/PhysRevD.83.024005}

\bibitem[{Gammie {et~al.}(2003)Gammie, McKinney, \& Toth}]{Gammie:2003rj}
Gammie, C.~F., McKinney, J.~C., \& Toth, G. 2003, Astrophys.J., 589, 444,
  \dodoi{10.1086/374594}

\bibitem[{{Gardiner} \& {Stone}(2005)}]{Gardiner:2005}
{Gardiner}, T.~A., \& {Stone}, J.~M. 2005, Journal of Computational Physics,
  205, 509, \dodoi{10.1016/j.jcp.2004.11.016}

\bibitem[{Gardiner \& Stone(2008)}]{Gardiner:2007nc}
Gardiner, T.~A., \& Stone, J.~M. 2008, J. Comput. Phys., 227, 4123,
  \dodoi{10.1016/j.jcp.2007.12.017}

\bibitem[{Giacomazzo \& Rezzolla(2007)}]{Giacomazzo:2007ti}
Giacomazzo, B., \& Rezzolla, L. 2007, Class. Quant. Grav., 24, S235,
  \dodoi{10.1088/0264-9381/24/12/S16}

\bibitem[{Goldberg {et~al.}(1967)Goldberg, MacFarlane, Newman, Rohrlich, \&
  Sudarshan}]{Goldberg:1966uu}
Goldberg, J.~N., MacFarlane, A.~J., Newman, E.~T., Rohrlich, F., \& Sudarshan,
  E. C.~G. 1967, J. Math. Phys., 8, 2155

\bibitem[{Goldstein {et~al.}(2017)}]{Goldstein:2017mmi}
Goldstein, A., {et~al.} 2017, Astrophys. J., 848, L14,
  \dodoi{10.3847/2041-8213/aa8f41}

\bibitem[{{Goodman}(1986)}]{Goodman:1986a}
{Goodman}, J. 1986, \apjl, 308, L47, \dodoi{10.1086/184741}

\bibitem[{Gourgoulhon {et~al.}(2001)Gourgoulhon, Grandclement, Taniguchi,
  Marck, \& Bonazzola}]{Gourgoulhon:2000nn}
Gourgoulhon, E., Grandclement, P., Taniguchi, K., Marck, J.-A., \& Bonazzola,
  S. 2001, Phys.Rev., D63, 064029, \dodoi{10.1103/PhysRevD.63.064029}

\bibitem[{Guercilena {et~al.}(2017)Guercilena, Radice, \&
  Rezzolla}]{Guercilena:2016fdl}
Guercilena, F., Radice, D., \& Rezzolla, L. 2017, Comput. Astrophys. Cosmol.,
  4, 3, \dodoi{10.1186/s40668-017-0022-0}

\bibitem[{Gundlach(1998)}]{Gundlach:1997us}
Gundlach, C. 1998, Phys. Rev. D, 57, 863, \dodoi{10.1103/PhysRevD.57.863}

\bibitem[{Gundlach \& Leveque(2011)}]{Gundlach:2010gy}
Gundlach, C., \& Leveque, R.~J. 2011, J. Fluid Mech., 676, 237,
  \dodoi{10.1017/jfm.2011.42}

\bibitem[{Hammond {et~al.}(2021)Hammond, Hawke, \& Andersson}]{Hammond:2021vtv}
Hammond, P., Hawke, I., \& Andersson, N. 2021, Phys. Rev. D, 104, 103006,
  \dodoi{10.1103/PhysRevD.104.103006}

\bibitem[{Hilditch {et~al.}(2013)Hilditch, Bernuzzi, Thierfelder, Cao, Tichy,
  \& Bruegmann}]{Hilditch:2012fp}
Hilditch, D., Bernuzzi, S., Thierfelder, M., {et~al.} 2013, Phys. Rev., D88,
  084057, \dodoi{10.1103/PhysRevD.88.084057}

\bibitem[{{Jiang} \& {Shu}(1996)}]{Jiang:1996a}
{Jiang}, G.-S., \& {Shu}, C.-W. 1996, Journal of Computational Physics, 126,
  202, \dodoi{10.1006/jcph.1996.0130}

\bibitem[{Kasen {et~al.}(2017)Kasen, Metzger, Barnes, Quataert, \&
  Ramirez-Ruiz}]{Kasen:2017sxr}
Kasen, D., Metzger, B., Barnes, J., Quataert, E., \& Ramirez-Ruiz, E. 2017,
  Nature, \dodoi{10.1038/nature24453}

\bibitem[{Kastaun(2006)}]{Kastaun:2006ik}
Kastaun, W. 2006, Phys. Rev. D, 74, 124024, \dodoi{10.1103/PhysRevD.74.124024}

\bibitem[{Kastaun {et~al.}(2021)Kastaun, Kalinani, \& Ciolfi}]{Kastaun:2020uxr}
Kastaun, W., Kalinani, J.~V., \& Ciolfi, R. 2021, Phys. Rev. D, 103, 023018,
  \dodoi{10.1103/PhysRevD.103.023018}

\bibitem[{Kidder {et~al.}(2017)}]{Kidder:2016hev}
Kidder, L.~E., {et~al.} 2017, J. Comput. Phys., 335, 84,
  \dodoi{10.1016/j.jcp.2016.12.059}

\bibitem[{Kiuchi {et~al.}(2015)Kiuchi, Cerdá-Durán, Kyutoku, Sekiguchi, \&
  Shibata}]{Kiuchi:2015sga}
Kiuchi, K., Cerdá-Durán, P., Kyutoku, K., Sekiguchi, Y., \& Shibata, M. 2015,
  Phys.\ Rev.\ D, 92, 124034, \dodoi{10.1103/PhysRevD.92.124034}

\bibitem[{Kiuchi {et~al.}(2022)Kiuchi, Fujibayashi, Hayashi, Kyutoku,
  Sekiguchi, \& Shibata}]{Kiuchi:2022nin}
Kiuchi, K., Fujibayashi, S., Hayashi, K., {et~al.} 2022.
\newblock \doarXiv{2211.07637}

\bibitem[{Kiuchi {et~al.}(2018)Kiuchi, Kyutoku, Sekiguchi, \&
  Shibata}]{Kiuchi:2017zzg}
Kiuchi, K., Kyutoku, K., Sekiguchi, Y., \& Shibata, M. 2018, Phys. Rev., D97,
  124039, \dodoi{10.1103/PhysRevD.97.124039}

\bibitem[{Kiuchi {et~al.}(2014)Kiuchi, Kyutoku, Sekiguchi, Shibata, \&
  Wada}]{Kiuchi:2014hja}
Kiuchi, K., Kyutoku, K., Sekiguchi, Y., Shibata, M., \& Wada, T. 2014,
  Phys.Rev., D90, 041502, \dodoi{10.1103/PhysRevD.90.041502}

\bibitem[{Kiuchi {et~al.}(2008)Kiuchi, Shibata, \& Yoshida}]{Kiuchi:2008ss}
Kiuchi, K., Shibata, M., \& Yoshida, S. 2008, Phys. Rev., D78, 024029,
  \dodoi{10.1103/PhysRevD.78.024029}

\bibitem[{{Komissarov}(1999)}]{Komissarov:1999a}
{Komissarov}, S.~S. 1999, Mon. Not. Roy. Astron. Soc., 303, 343,
  \dodoi{10.1046/j.1365-8711.1999.02244.x}

\bibitem[{Komissarov(2005)}]{Komissarov:2005wj}
Komissarov, S.~S. 2005, Mon. Not. Roy. Astron. Soc., 359, 801,
  \dodoi{10.1111/j.1365-2966.2005.08974.x}

\bibitem[{Kulkarni(2005)}]{Kulkarni:2005jw}
Kulkarni, S.~R. 2005.
\newblock \doarXiv{astro-ph/0510256}

\bibitem[{Kumar \& Zhang(2014)}]{Kumar:2014upa}
Kumar, P., \& Zhang, B. 2014, Phys. Rept., 561, 1,
  \dodoi{10.1016/j.physrep.2014.09.008}

\bibitem[{Lasky {et~al.}(2011)Lasky, Zink, Kokkotas, \&
  Glampedakis}]{Lasky:2011un}
Lasky, P.~D., Zink, B., Kokkotas, K.~D., \& Glampedakis, K. 2011, Astrophys. J.
  Lett., 735, L20, \dodoi{10.1088/2041-8205/735/1/L20}

\bibitem[{Li \& Paczynski(1998)}]{Li:1998bw}
Li, L.-X., \& Paczynski, B. 1998, Astrophys.J., 507, L59,
  \dodoi{10.1086/311680}

\bibitem[{Liebling {et~al.}(2010)Liebling, Lehner, Neilsen, \&
  Palenzuela}]{Liebling:2010bn}
Liebling, S.~L., Lehner, L., Neilsen, D., \& Palenzuela, C. 2010, Phys. Rev.,
  D81, 124023, \dodoi{10.1103/PhysRevD.81.124023}

\bibitem[{Liu {et~al.}(1994)Liu, Osher, \& Chan}]{Liu:1994}
Liu, X.-D., Osher, S., \& Chan, T. 1994, Journal of Computational Physics, 115,
  200 , \dodoi{http://dx.doi.org/10.1006/jcph.1994.1187}

\bibitem[{Liu {et~al.}(2008)Liu, Shapiro, Etienne, \& Taniguchi}]{Liu:2008xy}
Liu, Y.~T., Shapiro, S.~L., Etienne, Z.~B., \& Taniguchi, K. 2008, Phys. Rev.,
  D78, 024012, \dodoi{10.1103/PhysRevD.78.024012}

\bibitem[{L{\"o}ffler {et~al.}(2012)}]{Loffler:2011ay}
L{\"o}ffler, F., {et~al.} 2012, Class. Quant. Grav., 29, 115001,
  \dodoi{10.1088/0264-9381/29/11/115001}

\bibitem[{Londrillo \& Del~Zanna(2004)}]{Londrillo:2003qi}
Londrillo, P., \& Del~Zanna, L. 2004, J.Comput.Phys., 195, 17,
  \dodoi{10.1016/j.jcp.2003.09.016}

\bibitem[{{Markey} \& {Tayler}(1973)}]{Markey:1973a}
{Markey}, P., \& {Tayler}, R.~J. 1973, \mnras, 163, 77,
  \dodoi{10.1093/mnras/163.1.77}

\bibitem[{{Markey} \& {Tayler}(1974)}]{Markey:1974a}
---. 1974, \mnras, 168, 505, \dodoi{10.1093/mnras/168.3.505}

\bibitem[{Marti {et~al.}(1991)Marti, Ibanez, \& Miralles}]{Marti:1991wi}
Marti, J.~M., Ibanez, J.~M., \& Miralles, J.~A. 1991, Phys. Rev., D43, 3794,
  \dodoi{10.1103/PhysRevD.43.3794}

\bibitem[{{McCorquodale} {et~al.}(2015){McCorquodale}, {Dorr}, {Hittinger}, \&
  {Colella}}]{McCorquodale:2015a}
{McCorquodale}, P., {Dorr}, M.~R., {Hittinger}, J.~A.~F., \& {Colella}, P.
  2015, Journal of Computational Physics, 288, 181,
  \dodoi{10.1016/j.jcp.2015.01.006}

\bibitem[{Metzger {et~al.}(2010)Metzger, Martinez-Pinedo, Darbha, Quataert,
  Arcones, {et~al.}}]{Metzger:2010sy}
Metzger, B., Martinez-Pinedo, G., Darbha, S., {et~al.} 2010,
  Mon.Not.Roy.Astron.Soc., 406, 2650, \dodoi{10.1111/j.1365-2966.2010.16864.x}

\bibitem[{M{\"o}sta {et~al.}(2014)M{\"o}sta, Mundim, Faber, Haas, Noble,
  {et~al.}}]{Moesta:2013dna}
M{\"o}sta, P., Mundim, B.~C., Faber, J.~A., {et~al.} 2014, Class.Quant.Grav.,
  31, 015005, \dodoi{10.1088/0264-9381/31/1/015005}

\bibitem[{M\"osta {et~al.}(2020)M\"osta, Radice, Haas, Schnetter, \&
  Bernuzzi}]{Mosta:2020hlh}
M\"osta, P., Radice, D., Haas, R., Schnetter, E., \& Bernuzzi, S. 2020,
  Astrophys. J. Lett., 901, L37, \dodoi{10.3847/2041-8213/abb6ef}

\bibitem[{Narayan {et~al.}(1992)Narayan, Paczynski, \& Piran}]{Narayan:1992iy}
Narayan, R., Paczynski, B., \& Piran, T. 1992, Astrophys. J., 395, L83.
\newblock \doarXiv{astro-ph/9204001}

\bibitem[{Paczynski(1986)}]{Paczynski:1986px}
Paczynski, B. 1986, Astrophys. J., 308, L43

\bibitem[{Palenzuela {et~al.}(2022)Palenzuela, Aguilera-Miret, Carrasco,
  Ciolfi, Kalinani, Kastaun, Mi\~nano, \& Vigan\`o}]{Palenzuela:2021gdo}
Palenzuela, C., Aguilera-Miret, R., Carrasco, F., {et~al.} 2022, Phys. Rev. D,
  106, 023013, \dodoi{10.1103/PhysRevD.106.023013}

\bibitem[{Palenzuela {et~al.}(2018)Palenzuela, Mi\~nano, Vigan\`o, Arbona,
  Bona-Casas, Rigo, Bezares, Bona, \& Mass\'o}]{Palenzuela:2018sly}
Palenzuela, C., Mi\~nano, B., Vigan\`o, D., {et~al.} 2018, Class. Quant. Grav.,
  35, 185007, \dodoi{10.1088/1361-6382/aad7f6}

\bibitem[{Perego {et~al.}(2019)Perego, Bernuzzi, \& Radice}]{Perego:2019adq}
Perego, A., Bernuzzi, S., \& Radice, D. 2019, Eur. Phys. J., A55, 124,
  \dodoi{10.1140/epja/i2019-12810-7}

\bibitem[{Pian {et~al.}(2017)}]{Pian:2017gtc}
Pian, E., {et~al.} 2017, Nature, \dodoi{10.1038/nature24298}

\bibitem[{Pili {et~al.}(2014)Pili, Bucciantini, \& Del~Zanna}]{Pili:2014npa}
Pili, A.~G., Bucciantini, N., \& Del~Zanna, L. 2014, Mon. Not. Roy. Astron.
  Soc., 439, 3541, \dodoi{10.1093/mnras/stu215}

\bibitem[{Pili {et~al.}(2017)Pili, Bucciantini, \& Del~Zanna}]{Pili:2017yxd}
---. 2017, Mon. Not. Roy. Astron. Soc., 470, 2469,
  \dodoi{10.1093/mnras/stx1176}

\bibitem[{Piran(2004)}]{Piran:2004ba}
Piran, T. 2004, Rev. Mod. Phys., 76, 1143, \dodoi{10.1103/RevModPhys.76.1143}

\bibitem[{Plewa \& Mueller(1999)}]{Plewa:1998nma}
Plewa, T., \& Mueller, E. 1999, Astron. Astrophys., 342, 179.
\newblock \doarXiv{astro-ph/9807241}

\bibitem[{Pons \& Vigan\`o(2019)}]{Pons:2019zyc}
Pons, J.~A., \& Vigan\`o, D. 2019, \dodoi{10.1007/s41115-019-0006-7}

\bibitem[{Prakash {et~al.}(2021)Prakash, Radice, Logoteta, Perego, Nedora,
  Bombaci, Kashyap, Bernuzzi, \& Endrizzi}]{Prakash:2021wpz}
Prakash, A., Radice, D., Logoteta, D., {et~al.} 2021, Phys. Rev. D, 104,
  083029, \dodoi{10.1103/PhysRevD.104.083029}

\bibitem[{Pretorius(2005{\natexlab{a}})}]{Pretorius:2004jg}
Pretorius, F. 2005{\natexlab{a}}, Class.Quant.Grav., 22, 425,
  \dodoi{10.1088/0264-9381/22/2/014}

\bibitem[{Pretorius(2005{\natexlab{b}})}]{Pretorius:2005gq}
---. 2005{\natexlab{b}}, Phys. Rev. Lett., 95, 121101,
  \dodoi{10.1103/PhysRevLett.95.121101}

\bibitem[{Price \& Rosswog(2006)}]{Price:2006fi}
Price, D., \& Rosswog, S. 2006, Science, 312, 719,
  \dodoi{10.1126/science.1125201}

\bibitem[{Radice {et~al.}(2022)Radice, Bernuzzi, Perego, \&
  Haas}]{Radice:2021jtw}
Radice, D., Bernuzzi, S., Perego, A., \& Haas, R. 2022, Mon. Not. Roy. Astron.
  Soc., 512, 1499, \dodoi{10.1093/mnras/stac589}

\bibitem[{Radice {et~al.}(2018)Radice, Perego, Hotokezaka, Fromm, Bernuzzi, \&
  Roberts}]{Radice:2018pdn}
Radice, D., Perego, A., Hotokezaka, K., {et~al.} 2018, Astrophys. J., 869, 130,
  \dodoi{10.3847/1538-4357/aaf054}

\bibitem[{Radice {et~al.}(2014{\natexlab{a}})Radice, Rezzolla, \&
  Galeazzi}]{Radice:2013xpa}
Radice, D., Rezzolla, L., \& Galeazzi, F. 2014{\natexlab{a}},
  Class.Quant.Grav., 31, 075012, \dodoi{10.1088/0264-9381/31/7/075012}

\bibitem[{Radice {et~al.}(2014{\natexlab{b}})Radice, Rezzolla, \&
  Galeazzi}]{Radice:2013hxh}
---. 2014{\natexlab{b}}, Mon.Not.Roy.Astron.Soc., 437, L46,
  \dodoi{10.1093/mnrasl/slt137}

\bibitem[{{Rasio} \& {Shapiro}(1999)}]{Rasio:1999a}
{Rasio}, F.~A., \& {Shapiro}, S.~L. 1999, Classical and Quantum Gravity, 16,
  R1, \dodoi{10.1088/0264-9381/16/6/201}

\bibitem[{Reisswig {et~al.}(2013)Reisswig, Haas, Ott, Abdikamalov, Mösta,
  Pollney, \& Schnetter}]{Reisswig:2012nc}
Reisswig, C., Haas, R., Ott, C.~D., {et~al.} 2013, Phys. Rev., D87, 064023,
  \dodoi{10.1103/PhysRevD.87.064023}

\bibitem[{Rosswog \& Diener(2021)}]{Rosswog:2020kwm}
Rosswog, S., \& Diener, P. 2021, Class. Quant. Grav., 38, 115002,
  \dodoi{10.1088/1361-6382/abee65}

\bibitem[{Ruiz {et~al.}(2011)Ruiz, Hilditch, \& Bernuzzi}]{Ruiz:2010qj}
Ruiz, M., Hilditch, D., \& Bernuzzi, S. 2011, Phys. Rev., D83, 024025,
  \dodoi{10.1103/PhysRevD.83.024025}

\bibitem[{Ryu {et~al.}(1998)Ryu, Miniati, Jones, \& Frank}]{Ryu:1998ar}
Ryu, D., Miniati, F., Jones, T.~W., \& Frank, A. 1998, Astrophys. J., 509, 244,
  \dodoi{10.1086/306481}

\bibitem[{Savchenko {et~al.}(2017)}]{Savchenko:2017ffs}
Savchenko, V., {et~al.} 2017, Astrophys. J., 848, L15,
  \dodoi{10.3847/2041-8213/aa8f94}

\bibitem[{Shankar {et~al.}(2023)Shankar, M\"osta, Brandt, Haas, Schnetter, \&
  de~Graaf}]{Shankar:2022ful}
Shankar, S., M\"osta, P., Brandt, S.~R., {et~al.} 2023, Class. Quant. Grav.,
  40, 205009, \dodoi{10.1088/1361-6382/acf2d9}

\bibitem[{Shibata(1999)}]{Shibata:1999hn}
Shibata, M. 1999, Phys. Rev., D60, 104052, \dodoi{10.1103/PhysRevD.60.104052}

\bibitem[{Shibata \& Nakamura(1995)}]{Shibata:1995we}
Shibata, M., \& Nakamura, T. 1995, Phys. Rev., D52, 5428,
  \dodoi{10.1103/PhysRevD.52.5428}

\bibitem[{Shibata \& Sekiguchi(2005)}]{Shibata:2005gp}
Shibata, M., \& Sekiguchi, Y.-i. 2005, Phys. Rev., D72, 044014,
  \dodoi{10.1103/PhysRevD.72.044014}

\bibitem[{Shibata \& Uryu(2000)}]{Shibata:1999wm}
Shibata, M., \& Uryu, K. 2000, Phys. Rev., D61, 064001,
  \dodoi{10.1103/PhysRevD.61.064001}

\bibitem[{Siegel {et~al.}(2014)Siegel, Ciolfi, \& Rezzolla}]{Siegel:2014ita}
Siegel, D.~M., Ciolfi, R., \& Rezzolla, L. 2014, Astrophys. J., 785, L6,
  \dodoi{10.1088/2041-8205/785/1/L6}

\bibitem[{Siegel \& Metzger(2017)}]{Siegel:2017nub}
Siegel, D.~M., \& Metzger, B.~D. 2017, Phys. Rev. Lett., 119, 231102,
  \dodoi{10.1103/PhysRevLett.119.231102}

\bibitem[{Siegel {et~al.}(2018)Siegel, M\"osta, Desai, \& Wu}]{Siegel:2017sav}
Siegel, D.~M., M\"osta, P., Desai, D., \& Wu, S. 2018, Astrophys. J., 859, 71,
  \dodoi{10.3847/1538-4357/aabcc5}

\bibitem[{Soares-Santos {et~al.}(2017)}]{Soares-santos:2017lru}
Soares-Santos, M., {et~al.} 2017, Astrophys. J., 848, L16,
  \dodoi{10.3847/2041-8213/aa9059}

\bibitem[{Steiner {et~al.}(2013)Steiner, Hempel, \& Fischer}]{Steiner:2012rk}
Steiner, A.~W., Hempel, M., \& Fischer, T. 2013, Astrophys. J., 774, 17,
  \dodoi{10.1088/0004-637X/774/1/17}

\bibitem[{Stergioulas \& Friedman(1995)}]{Stergioulas:1994ea}
Stergioulas, N., \& Friedman, J.~L. 1995, Astrophys. J., 444, 306,
  \dodoi{10.1086/175605}

\bibitem[{Stone {et~al.}(2008)Stone, Gardiner, Teuben, Hawley, \&
  Simon}]{Stone:2008mh}
Stone, J.~M., Gardiner, T.~A., Teuben, P., Hawley, J.~F., \& Simon, J.~B. 2008,
  Astrophys. J. Suppl., 178, 137, \dodoi{10.1086/588755}

\bibitem[{{Stone} {et~al.}(2020){Stone}, {Tomida}, {White}, \&
  {Felker}}]{Stone:2020}
{Stone}, J.~M., {Tomida}, K., {White}, C.~J., \& {Felker}, K.~G. 2020,
  Astrophys.\ J.\ Suppl., 249, 4, \dodoi{10.3847/1538-4365/ab929b}

\bibitem[{Sur {et~al.}(2022)Sur, Cook, Radice, Haskell, \&
  Bernuzzi}]{Sur:2021awe}
Sur, A., Cook, W., Radice, D., Haskell, B., \& Bernuzzi, S. 2022, Mon. Not.
  Roy. Astron. Soc., 511, 3983, \dodoi{10.1093/mnras/stac353}

\bibitem[{{Tayler}(1957)}]{Tayler:1957a}
{Tayler}, R.~J. 1957, Proceedings of the Physical Society B, 70, 31,
  \dodoi{10.1088/0370-1301/70/1/306}

\bibitem[{{Tayler}(1973)}]{Tayler:1973a}
---. 1973, \mnras, 161, 365, \dodoi{10.1093/mnras/161.4.365}

\bibitem[{Thierfelder {et~al.}(2011{\natexlab{a}})Thierfelder, Bernuzzi, \&
  Br{\"u}gmann}]{Thierfelder:2011yi}
Thierfelder, M., Bernuzzi, S., \& Br{\"u}gmann, B. 2011{\natexlab{a}},
  Phys.Rev., D84, 044012, \dodoi{10.1103/PhysRevD.84.044012}

\bibitem[{Thierfelder {et~al.}(2011{\natexlab{b}})Thierfelder, Bernuzzi,
  Hilditch, Br{\"u}gmann, \& Rezzolla}]{Thierfelder:2010dv}
Thierfelder, M., Bernuzzi, S., Hilditch, D., Br{\"u}gmann, B., \& Rezzolla, L.
  2011{\natexlab{b}}, Phys.Rev., D83, 064022,
  \dodoi{10.1103/PhysRevD.83.064022}

\bibitem[{{T{\'o}th} \& {Roe}(2002)}]{Toth:2002a}
{T{\'o}th}, G., \& {Roe}, P.~L. 2002, Journal of Computational Physics, 180,
  736, \dodoi{10.1006/jcph.2002.7120}

\bibitem[{Typel {et~al.}(2015)Typel, Oertel, \& Kl\"ahn}]{Typel:2013rza}
Typel, S., Oertel, M., \& Kl\"ahn, T. 2015, Phys. Part. Nucl., 46, 633,
  \dodoi{10.1134/S1063779615040061}

\bibitem[{Tóth(2000)}]{Toth:2000}
Tóth, G. 2000, Journal of Computational Physics, 161, 605 ,
  \dodoi{10.1006/jcph.2000.6519}

\bibitem[{{van Leer}(1974)}]{vanLeer:1974a}
{van Leer}, B. 1974, Journal of Computational Physics, 14, 361,
  \dodoi{10.1016/0021-9991(74)90019-9}

\bibitem[{Vigan\`o {et~al.}(2018)Vigan\`o, Mart\'\i{}nez-G\'omez, Pons,
  Palenzuela, Carrasco, Mi\~nano, Arbona, Bona, \& Mass\'o}]{Vigano:2018lrv}
Vigan\`o, D., Mart\'\i{}nez-G\'omez, D., Pons, J.~A., {et~al.} 2018,
  \dodoi{10.1016/j.cpc.2018.11.022}

\bibitem[{Weyhausen {et~al.}(2012)Weyhausen, Bernuzzi, \&
  Hilditch}]{Weyhausen:2011cg}
Weyhausen, A., Bernuzzi, S., \& Hilditch, D. 2012, Phys. Rev., D85, 024038,
  \dodoi{10.1103/PhysRevD.85.024038}

\bibitem[{White {et~al.}(2023)White, Mullen, Jiang, Davis, Stone, Morozova, \&
  Zhang}]{White:2023wxh}
White, C.~J., Mullen, P.~D., Jiang, Y.-F., {et~al.} 2023, Astrophys. J., 949,
  103, \dodoi{10.3847/1538-4357/acc8cf}

\bibitem[{White {et~al.}(2016)White, Stone, \& Gammie}]{White:2015omx}
White, C.~J., Stone, J.~M., \& Gammie, C.~F. 2016, Astrophys. J. Suppl., 225,
  22, \dodoi{10.3847/0067-0049/225/2/22}

\bibitem[{{Wright}(1973)}]{Wright:1973a}
{Wright}, G.~A.~E. 1973, \mnras, 162, 339, \dodoi{10.1093/mnras/162.4.339}

\end{thebibliography}
\end{document}